\pdfoutput=1
\documentclass[12pt,a4paper]{article}
\usepackage{amsmath}
\usepackage{amssymb}
\usepackage{graphicx}
\usepackage{verbatim}
\usepackage{ifthen}
\usepackage{epsfig}
\usepackage{setspace}
\usepackage[left=2cm,right=2cm,top=2cm,bottom=2cm]{geometry}

\newcounter{fig}   \newcommand{\lbfig}[1]{\refstepcounter{fig}
\label{#1} }

\newcommand{\bea}{\begin{eqnarray}}
\newcommand{\eea}{\end{eqnarray}}
\newcommand{\be}{\begin{equation}}
\newcommand{\ee}{\end{equation}}

\def\bfph{{\pmb{\phi}}}

\newcommand{\re}[1]{(\ref{#1})}

\newcommand{\CL}{\ensuremath{\mathcal{L}}}

\newcommand{\pa}{\partial}

\usepackage{graphicx}
\usepackage{graphpap,geometry}
\usepackage[outdir=./]{epstopdf}
\usepackage{ulem}

\title{Gauged Baby Skyrme Model with Chern-Simons term}
\author{
{\large A.~Samoilenka}$^{\dagger}$
and {\large Ya. Shnir}$^{\dagger \star}$
 \\ \\
\\ $^{\dagger}${\small Department of Theoretical Physics and Astrophysics}\\
{\small Belarusian State University, Minsk 220004, Belarus}
\\ $^{\star}${\small BLTP, JINR, Dubna, Russia}
}

\begin{document}

\maketitle

\begin{abstract}
The properties of the multisoliton solutions of the (2+1)-dimensional Maxwell-Chern-Simons-Skyrme model
are investigated numerically. Coupling to the Chern-Simons term allows for existence of the electrically charge solitons which may also carry
magnetic fluxes. Two  particular choices of the potential term is considered:
(i) the weakly bounded potential and (ii) the double vacuum potential. In the absence of the gauge interaction in the former case
the individual constituents of the multisoliton configuration are well separated, while in the latter case the rotational invariance of the
configuration remains unbroken.
It is shown that coupling of the planar multi-Skyrmions to the electric and magnetic field
strongly affects the pattern of interaction between the constituents. We analyze the dependency of the
structure of the solutions, the energies, angular momenta, electric and  magnetic fields of the configurations
on the gauge coupling constant $g$, and the electric potential.
It is found that, generically, the coupling to the Chern-Simons term strongly affects the usual pattern of interaction
between the skyrmions, in particular the electric repulsion between the solitons may break the multisoliton configuration into
partons. We show that as the gauge coupling becomes strong, both the magnetic flux and the electric charge of the solutions become
quantized although they are not topological numbers.
\end{abstract}
\maketitle
%%%%%%%%%%%%%%%%%%%%%%%%%%%%%%%%%%%%%%%%%%%%%%%%%%%%%%%%%%%%%%%%%%
\section{Introduction}
%%%%%%%%%%%%%%%%%%%%%%%%%%%%%%%%%%%%%%%%%%%%%%%%%%%%%%%%%%%%%%%%%%
Diverse models of the field theory support regular soliton solutions. Many such
models have been intensively studied over last decades in a wide variety of
physical contexts. Perhaps one of the most interesting examples is (2+1)-dimensional
non-linear $O(3)$ sigma model, which is also known as the baby Skyrme model.

Soliton solutions of this model were considered for the first time in the papers
\cite{BB,Leese:1989gj}. A few years later, they were revisited in
\cite{Bsk} the term "Baby Skyrmion" was coined in the second of these paper to
describe planar reduction of the original Skyrme model. Indeed, this low-dimensional simplified non-linear theory emulates the
conventional Skyrme model in $\left( 3+1\right) $ dimensions \cite{Skyrme:1961vq} in many respects. In particular,
this planar theory was used to study the processes of scattering
of the solitons \cite{PZS,Leese:1989gj} and the effects of isorotations of the multisolitons
\cite{Battye:2013tka,Halavanau:2013vsa}. The low dimensionality of the model significantly simplifies full numerical
computations making it possible to exclude any restrictions imposed by the parametrization of the fields.

A peculiar feature of the planar Skyrme model is that the structure of the
multisoliton configurations is very sensitive to the particular choice of the potential of the model
\cite{Ward,Hen,JSS,Salmi:2014hsa}.
Further, a suitable choice for the potential term allows us to
split the individual constituents of the planar Skyrmions, each
of them being associated with a fractional part of the topological
charge of the configuration \cite{JSS}. Another choice of the potential term allows us to
construct weakly bounded multisoliton configurations via
combination of a short-range repulsion and a long-range attraction
between the solitons \cite{Salmi:2014hsa}.

There has recently been significant progress in construction of various multisoliton configurations in the baby Skyrme model
\cite{Bsk,Ward,Hen,JSS,Salmi:2014hsa}. Also it was found that the restricted baby Skyrme model in (2+1)-dimensions
\cite{Gisiger:1995yb,Gisiger:1996vb,Arthur:1996ia} supports BPS solutions saturating the topological bound.
Furthermore, the BPS Skyrme model is integrable in the sense of
generalized integrability, it supports an infinite number of conservation laws \cite{Adam:2010jr}.
The planar Skyrme model in AdS spacetime was also considered as a
low-dimensional analogue of Sakai-Sugimoto model of holographic QCD \cite{Bolognesi:2013jba}.

The baby Skyrme model attracts a special attention since this simple theory finds various direct physical realizations.
Indeed, this model originally was formulated as a modification of the Heisenberg-type models of
interacting spins \cite{BB}.
Further, hexagonal lattices of two-dimensional skyrmions were observed in a thin
ferromagnetic layer \cite{Yu}, and in a metallic itinerant-electron
magnet, where the Skyrmion lattice was detected by results of neutron
scattering \cite{Muhlenbauer}.
The Skyrmion configurations naturally arise in various condensed matter systems with intrinsic and induced chirality,
some modification of the baby Skyrme model with the Dzyaloshinskii-Moriya
interaction term was suggested to model noncentrosymmetric ferromagnetic planar structures \cite{Bogdanov}.
These solitonic states were considered in the context of future applications in development of data storage technologies
and emerging spintronics, see e.g. \cite{Heinze}.

Planar Skyrmions yield a specific contribution in the description of the
topological quantum Hall effect \cite{Girvin,Neubauer}. In this framework the Skyrmion-like states are coupled to fluxes of magnetic
field and carry an electric charge.
Therefore, in order to study the topological properties of planar quantum systems,
it is crucial to extend the low-dimensional Skyrme model by gauging it and coupling to the electric and magnetic fields.

The planar Skyrme-Maxwell model was considered in \cite{Gladikowski:1995sc,Samoilenka:2015bsf}. An interesting observation is that
in the strong coupling regime the magnetic flux coupled to the Skyrmion, is quantized thought there is no
topological reason of that. Further, it is found recently that, in the model with weakly bounding potential,
the coupling to the magnetic field results in effective recovering of the rotational invariance of the
configuration \cite{Samoilenka:2015bsf}.

It is known that the electric field is always trivial in the planar gauged Skyrme model \cite{Gladikowski:1995sc}. In order to
construct electrically charged planar Skyrmions one has to include the
Chern-Simons term  in the Lagrangian of the model \cite{Loginov}. Then the solitons may acquire electric charge in addition to the
magnetic flux. Thus, we can expect the general picture of interaction between the Skyrmions  becomes more complicated and various pattern
of symmetry breaking may be observed.

The key ingredient in the discussion of the properties of the soliton solutions of the planar
Maxwell-Chern-Simons-Skyrme model presented recently in \cite{Loginov} is the assumption of rotational invariance of the solutions.
Indeed, it simplifies the consideration significantly since the problem then can be reduced to the numerical
solution of system of three coupled ordinary differential equations. However, the rotational invariance is not a general property of
the multisoliton solution of this model, thus this problem should be revisiting.

In this paper we investigate the Maxwell-Chern-Simons-Skyrme model in (2+1) dimensions and study the
resulting multisoliton solutions. Since the choice of potential is critical for the structure of the solitons of the baby Skyrme model,
we consider the weakly bounded potential \cite{Salmi:2014hsa} and
the double vacuum (or "easy-axis" potential) \cite{BB,Weidig:1998ii}. This choice is motivated by the fact that in the case of zero gauge
coupling the individual constituents of the multisoliton configuration in the model with weakly bounded potential
are well separated, while in the model with double vacuum potential the rotational invariance of the configuration always unbroken.

Our calculations are performed for multisoliton solutions
up to charge $Q=5$ without any restrictions of symmetry. We study numerically the dependency of the
structure of the solutions, their energies, angular momenta,  electric and  magnetic fields
on the gauge coupling constant $g$, and electric potential given by the value of $A_0=\omega$ at spatial infinity.

\section{Model}
We consider a gauged version of the $O(3)$ $\sigma$-model with the Skyrme term in $2+1$ dimensions
with a Lagrangian \cite{Loginov}
\be
\label{lag}
L = -\frac{1}{4}F_{\mu\nu}F^{\mu\nu} +\frac{c}{4}\varepsilon^{\mu\nu\rho}F_{\mu\nu}A_\rho +
\frac{1}{2}D_\mu \vec\phi \cdot D^\mu \vec\phi - \frac{1}{4}(D_\mu \vec\phi \times D_\nu \vec\phi)^2 - V(\vec\phi)
\ee
Here the triplet of scalar fields  $\vec \phi$ is constrained to the surface of a sphere of unit radius:
$\vec \phi \cdot \vec \phi=1$. We
introduced the usual Maxwell term with the field strength tensor defined as
$F_{\mu\nu}=\partial_\mu A_\nu -\partial_\nu A_\mu$. The flat metric is
$g_{\mu\nu}=diag(1,-1,-1)$ and the coupling of the Skyrme field to the magnetic field is
given by the covariant derivative \cite{Adam:2012pm,Schroers:1995he,Gladikowski:1995sc}
$$
D_\mu \vec{\phi}=\pa_\mu \vec{\phi}+gA_\mu \vec{\phi}\times\vec{\phi}_\infty
$$

Topological restriction on the field $\vec \phi$ is that it approaches the
vacuum at spacial boundary, i.e. $\vec \phi_\infty = (0,0,1)$.
Thus, the field $\phi^a$ of the finite energy regular baby Skyrmion configuration is
a map $\bfph:\mathbb{R}^2 \to \mathbb{S}^2$ which belongs
to an equivalence class characterized by the topological charge $Q = \pi_2(\mathbb{S}^2) = \mathbb{Z}$.
Explicitly,
\be \label{charge}
Q= - \frac{1}{4\pi}\int \vec{\phi} \cdot   (\partial_1 \vec{\phi} \times \partial_2 \vec{\phi}) ~d^2x
\ee
Further, we suppose that the potential $V(\vec\phi)$ breaks the symmetry to $SO(2) \simeq U(1)$. This unbroken
subgroup corresponds to the $U(1)$ gauge transformations of the fields
\be
\label{gauge}
A^\prime_\mu=A_\mu + \frac{i}{g} U(\alpha) \partial_\mu U(\alpha)^{-1};\qquad \vec \phi_\perp^\prime =
U(\alpha) \vec \phi_\perp
\ee
where  $U(\alpha) = e^{ig \alpha(x)}$ and $\vec \phi_\perp = (\phi_1 + i\phi_2)$. Thus, the third component of the
Skyrme field $\phi_3$ remains decoupled. However since the fields are restricted to the surface of the unit sphere, coupling
of the planar components $\vec\phi_\perp$ to the gauge sector affects the component $\phi_3$ indirectly.

Note that the transformations of the Skyrme field $\vec \phi$ are actually isototations around the third axis in the
internal space, which (in the absence of the gauge field)
were considered in \cite{Battye:2013tka,Halavanau:2013vsa}.

The Chern-Simons term $L_{CS} = \frac{c}{4}\varepsilon^{\mu\nu\rho}F_{\mu\nu}A_\rho$ is topological
\cite{Jackiw:1980kv,Schonfeld:1980kb,Deser:1982vy,Paul:1986ix}. Under the gauge transformations \re{gauge}
it changes as total derivative,
$L_{CS} \to L_{CS} + \frac{c}{2} \partial_\mu (\alpha \varepsilon^{\mu\nu\rho} \partial_\nu A_\rho)$.
Thus, if the boundary terms can be neglected, the corresponding action of the
Maxwell-Chern-Simons-Skyrme model is invariant with respect to the gauge transformations \re{gauge}.
Note that the Chern-Simons term violates $P$ and $T$ invariance of the system \re{lag} while it
preserves $C$ symmetry. Consequently, the left and right isorotations of the Skyrme field \re{gauge}
are not identical.

Coupling of the Skyrme field to the Chern-Simons term  has many interesting consequences \cite{Arthur:1996uu}. Note that since this
term is independent of the metric,  it does not contribute to the
energy-momentum tensor of the system, which follows from \re{lag}:
\be
\begin{split}
T_{\mu\nu}=&-F_{\mu\lambda}F_\nu^{\ \lambda}
+ \frac{1}{4}g_{\mu\nu}F_{\lambda\rho}F^{\lambda\rho}+D_\mu\vec{\phi}\cdot D_\nu\vec{\phi} -
(D_\mu\vec{\phi}\times D_\rho\vec{\phi})\cdot( D_\nu\vec{\phi}\times D^\rho\vec{\phi})\\
&-
g_{\mu\nu}\left[\frac{1}{2}D_\rho\vec{\phi}\cdot D^\rho\vec{\phi} -
\frac{1}{4}(D_\rho\vec{\phi}\times D_\sigma\vec{\phi})\cdot( D^\rho\vec{\phi}\times D^\sigma\vec{\phi}) -
V \right]
\end{split}
\ee
Thus, the energy  of the system is the sum of the kinetic energy
\be
\label{kin}
T = \frac12 \int \left[E_{i}^2 + (D_0\vec \phi)^2 + (D_0 \vec\phi \times D_i \vec\phi)^2 \right] ~d^2x
\ee
and the potential energy
\be
W=\frac12 \int \left[B^2 + (D_1\vec \phi)^2 + (D_2\vec \phi)^2 + (D_1 \vec\phi \times D_2 \vec\phi)^2
+ V(\vec\phi) \right]~d^2x
\ee
where the electric components of the field strength are $E_i=F_{0i}$ and the magnetic component is $B=F_{12}$.
Integration over all 2-dimensional space of the energy density $T_{00}$ yields the total mass-energy of the system
$E=\int T_{00}~d^2x$.

The complete set of the field equations, which follows from the variation of the action of the
model \re{lag}, is \cite{Gladikowski:1995sc,Loginov}
\be
\label{eqs}
\begin{split}
&D_\mu \vec J^\mu =\frac{\pa V}{\pa \vec \phi}\times \vec \phi \,  ;  \\
&\partial_\mu F^{\mu\nu}+
\frac{c}{2}\varepsilon^{\nu\alpha\beta}F_{\alpha\beta} = g\vec \phi_\infty \cdot \vec J^\nu \, ,
\end{split}
\ee
where \cite{Gladikowski:1995sc}
\be
\vec J^\mu=\vec \phi \times D^\mu \vec \phi - D_\nu \vec \phi  (D^\nu \vec \phi \cdot \vec \phi \times D^\mu \vec \phi )
\ee
Thus, the conserved current is $j_\mu = \vec \phi_\infty \cdot \vec J_\mu$ and this current is a source for the
electromagnetic field.  Note that the Gauss law from the variation of $A_0$ is $\nu=0$ component of the second equation
\re{eqs}:
\be
\label{Gauss}
\partial_i E_i +c B = g j_0
\ee
Another effect of the Chern-Simons term, which is evident from the second equation \re{eqs},  is that the gauge field
becomes massive, the mass of the photon is given by the constant $c$ \cite{Deser:1982vy}.

Notably, it is known that in the Chern-Simons-Maxwell theory the total magnetic flux through the $x-y$-plane, $\Phi = \int d^2 x B$,
is proportional to the electric charge, $q=c \Phi$  \cite{Jackiw:1980kv,Schonfeld:1980kb,Deser:1982vy}.
Unlike the Abelian Higgs model with Chern-Simons term \cite{deVega:1986eu},
in the planar Skyrme theory the magnetic flux is in general, non-quantized. However, in the regime of strong gauge coupling,
the magnetic flux becomes quantized without any topological reason \cite{Gladikowski:1995sc,Samoilenka:2015bsf}.
Note that since we do not restrict the symmetry of the configuration,
the integral relation between the total electric charge  and total magnetic flux of the system
does not impose any restriction on the electric charges and magnetic fluxes of the constituents.

We consider electromagnetic field generated by the Maxwell potential
\be
\label{el-ansatz}
A_0=A_0(x,y);\qquad A_1=A(x,y)\, ;\qquad A_2 = 0;
\ee
where  the gauge fixing condition is
used to exclude the $A_2$ component of the vector-potential. Thus the magnetic field is orthogonal to the $x-y$ plane:
$B = (0,0,B) = -\partial_2 A_1$ while the electric field in restricted to this plane,
$E_i=-\pa_i A_0$.  From the condition of finiteness of the energy at the spacial boundary we have to assume
that both the electric and magnetic fields are vanishing as $r \to \infty$. However,
the electric potential $A_0\rightarrow ~\omega$ at spacial infinity, where the constant $\omega$ is a free parameter which
yields the total electric charge of the configuration.

Now we can make use of the remaining gauge degree of freedom, which corresponds to the time-dependent
gauge transformations $U(t)=e^{i g \chi(t)}$:
$$
A_0 \to A_0^\prime = A_0 +\frac{i}{g} U(t) \partial_t U(t)^{-1}, \qquad D_0\vec \phi \to U(t) D_0 \vec \phi \, .
$$
Particular choice of the gauge function $\chi(t) = -\omega t$ yields $A_0^\prime = A_0 -\omega$, therefore
the boundary condition on the gauge transformed electric potential is $A_0^\prime \rightarrow 0$ as $x,y \to \infty$.
In other words, we can write the kinetic energy of the system \re{kin} as
\be
T = \frac12 \int \left[E_{i}^2 + g^2 (A_0^\prime + \omega)^2
[(\phi_1)^2+(\phi_2)^2+(\partial_i\phi_3)^2] \right] ~d^2x
\ee
Note that the term $\sim g^2(A_0^\prime + \omega)^2[(\phi_1)^2+(\phi_2)^2]$ in the total Hamiltonian of the system effectively
contributes to the mass of the corresponding components of scalar field. As we will see it strongly affects the properties of multisoliton
configurations.

For the sake of compactness of notations we omit the superscript '$^\prime$'
henceforth and suppose that the electromagnetic potential
$A_0$ is vanishing at the boundary. Then the Chern-Simons term in the Lagrangian \re{lag} can be written as
\be
\CL_{CS}=\int{ \left\{ \frac{c}{2}(\pa_2 A_0 A_1-\pa_2 A_1 A_0) \right\}}d^2x=
\left\{up\ to\ total\ derivative\right\}=-c\int{ A_0\pa_2 A_1}d^2x \, .
\ee

An important property of the electrically charged solitions is that they usually possess an intrinsic angular momentum, associated
with internal rotation of the configuration. The total angular momentum of the system \re{lag} is given by
\be
\label{angular}
J=\int T_{\varphi 0}~d^2x \, , \qquad \textrm{where}\qquad  T_{\varphi 0}=x T_{2 0}-y T_{1 0}
\ee
with
\be
\begin{split}
T_{1 0}&=\pa_2 A_0\pa_2A_1 + g A_0 \left[ (\vec\phi\times\vec{\phi}_\infty)\cdot
D_1\vec\phi - \pa_2\phi_3 (\vec\phi\cdot(D_1\vec\phi\times D_2\vec\phi)) \right]\, ;\\
T_{2 0}&=-\pa_1 A_0\pa_2A_1 + g A_0 \left[ (\vec\phi\times\vec{\phi}_\infty)
\cdot D_2\vec\phi + \pa_1\phi_3 (\vec\phi\cdot(D_1\vec\phi\times D_2\vec\phi)) \right]
\end{split}
\ee

Note that usually there is a linear relation $J\sim q$ between the
angular momentum of the spinning solitons and its electric charge. This
relation holds for Q-balls \cite{Coleman:1985ki,Volkov:2003ew}, monopole-antimonopole pairs \cite{VanderBij:2001nm,Paturyan:2004ps}
and gauged Skyrmions in (3+1)-dimensions \cite{Radu:2005jp}. However, in the presence of the Chern-Simons term, this relation is broken
since this term violates the spacial parity but preserves the charge conjugation. For example,
in the sector of degree one more complicated relation holds \cite{Loginov}
\be
J=-\frac{q}{g} + \frac{q^2}{4\pi c}\,  .
\ee
In sectors of higher degree the situation becomes more involved unless no restriction of rotational symmetry is
imposed, thus its generalization suggested in \cite{Loginov} turns out to be an
artefact of the hedgehog approximation.

%%%%%%%%%%%%%%%%%%%%%%%%%%%%%%%%%%
\section{Soliton solutions}
%%%%%%%%%%%%%%%%%%%%%%%%%%%%%%%%%%

In this section we investigate static soliton solutions of the model \re{lag}.
Generally, when a nonvanishing electric field is presented in a system, the functional whose stationary points we are looking for,
is not the positively defined energy but the indefinite action \cite{Yang}. Further, since
the Chern-Simons term does not contribute to the
energy functional and it appears in the action as a boundary term, the
usual approach to this problem is to look for corresponding solutions of the Euler-Lagrange equations
which follow from the Lagrangian \re{lag} subject to a set of the boundary conditions \cite{Loginov}.
We implemented this approach in order to  check of the correctness of our results, which were obtained in a different way.

Indeed, the standard energy minimization technique cannot be applied in this case, but we can consider two related
static functionals
\be
\label{ham}
\begin{split}
\mathcal{H}_1=&\int \biggl\{ \frac{1}{2}(\pa_2 A_1+c A_0)^2 + \frac{g^2 (A_0+\omega)^2}{2}\left[(\phi_3)^2-1-(\pa_i \phi_3)^2\right]
+ \frac{1}{2}D_{i}\vec{\phi}\cdot D_{i}\vec{\phi}\\
&+ \frac{1}{4}\left(D_i \vec{\phi}\times D_j \vec{\phi}\right)^2 + V \biggr\}~d^2x
\end{split}
\ee
and
\be
\label{eps0}
\mathcal{H}_2=\int \left\{ \frac{1}{2}(\pa_1 A_0)^2 + \frac{1}{2}(\pa_2 A_0 +c A_1)^2
- \frac{1}{2}(c A_1)^2 + \frac{g^2 (A_0+\omega)^2}{2}\left[1-(\phi_3)^2+(\pa_i \phi_3)^2\right] \right\}~d^2x
\ee

To construct multisoliton solutions of the model \re{lag} we simultaneously minimize the functional \re{ham}
with respect to variables $A_1$ and $\vec\phi$ leaving $A_0$ constant, and the second functional \re{eps0},
with respect to $A_0$ leaving $A_1$ and $\vec\phi$ constant. However, in our
calculations we do not adopt any \textit{a priori }assumptions about spatial
symmetries of the fields, both in the gauge and in the scalar sectors.

One can see that the stationary points of both quantities correspond to the same set of the
field equations \re{eqs} because, up to the terms, which do not affect the equations of motion,
we have $\mathcal{H}_1=-\CL$ and $\mathcal{H}_2=\CL$.

The numerical calculations are mainly performed on a equidistant square grid,
typically containing $200^2$  lattice points and with a lattice spacing $dx=0.15$
To check our results for consistency we also considered the lattice spacings $dx=0.1,0.2$.

The numerical algorithm employed was similar to that used in \cite{Samoilenka:2015bsf,Hale:2000fk}.
Well-chosen initial configurations of given degree produced via rational map ansatz were evolved
using the Metropolis method to minimize the functionals \re{ham},\re{eps0}.

Let us now briefly discuss the parameters of the model \re{lag}. Each configuration in a sector
of given degree $Q$ is labeled by three parameters, the gauge coupling constant $g$,
Chern-Simons couplng $c$ and the value of the electric potential $\omega$.

We also are free to choose the explicit form of the potential term $V$,
the structure of multisoliton configurations strongly depends on it. In what follows, we consider two possibilities,
so called double vacuum potential \cite{BB,Weidig:1998ii}
\be \label{double}
V = \mu^2 (1-\phi_3^2)\, ,
\ee
and lightly bound planar model with the potential \cite{Salmi:2014hsa,Samoilenka:2015bsf}
\be \label{Salmi}
V = \mu^2\left[\lambda(1 - \phi_3) + (1-\lambda)(1 - \phi_3)^4\right] \, , \quad \lambda \in [0,1] \, ,
\ee
If we restrict our consideration to the case $\lambda=0.5$ this potential
combines a short-range repulsion and a long-range attraction between the solitons, thus the
multisoliton configuration consists of well
separated partons with small binding energies.

The parameter $\mu^2$ in both cases corresponds to the mass of the scalar field. In the case of the double vacuum potential
the interaction between the solitons is strongly attractive, the rotational invariance of the multisoliton configurations is unbroken and
the isorotation of the system almost does not violate it \cite{Battye:2013tka,Halavanau:2013vsa}.

Note that the choice of the electric potential $\omega$ is restricted by the requirement of finiteness of energy \cite{Loginov}
\be
\label{restr-pot}
 |\omega|<\frac{1}{g}\sqrt{-\frac{\pa V}{\pa \phi_3}\bigg|_{\phi_3\rightarrow1}}
\ee
Actually this restriction appears on the same physical reasons as the restriction on the
angular frequency of the isospinning baby Skyrmions
\cite{Battye:2013tka,Halavanau:2013vsa}, the mass term is necessary to stabilize the soliton.

%%%%%%%%%%%%%%%%%%%%%%%%%%%%%%%%%%%%%%%
\section{Numerical results}
%%%%%%%%%%%%%%%%%%%%%%%%%%%%%%%%%%%%%%%
\subsection{Weakly bound Skyrmions}

Let us first consider the potential \re{Salmi}. In this case the electric potential is restricted as
\be
|\omega|<\frac{1}{g}\sqrt{-\frac{\pa V}{\pa \phi_3}\bigg|_{\phi_3\rightarrow1}}=\frac{\sqrt{\lambda \mu^2}}{g}
\ee
In most of our calculations we choose $\mu^2=0.1$, $\lambda=0.5$ and set $c=1$. Our goal is to analyse how the structure of the
multisoliton configurations will be affected by the variation of the rescaled parameter
$\omega \rightarrow\frac{\sqrt{\lambda \mu^2}}{g}\omega$, thus $\omega \in [-1,1]$.
In a sector of given topological degree $Q$ each static
configuration is characterized by its energy $E$, total electromagnetic energy $E_{em}$ and
distributions of the electric field $\vec E$, the magnetic flux $\Phi$ and the angular momentum $J$.

First, we set  $\omega=0$ and study the dependence of these quantities on the strength of the  gauge coupling
$g \in [0,2]$, see Fig.~\ref{overg}.
As the gauge coupling increases from zero, the energy of the
configuration decreases since the magnetic flux is
formed, as shown in the left upper plot of Fig.~\ref{overg}.
Due to the Chern-Simons coupling, also the electric field is generated
by some distribution of the electric charge density on the $x-y$ plane.

The distribution of the magnetic flux is associated with the position of the solitons, it
is orthogonal to the $x-y$ plane \cite{Gladikowski:1995sc},
the fluxes are attached to the individual partons of the
multi-soliton configuration.

As in the case of the model without the Chern-Simons term \cite{Gladikowski:1995sc,Samoilenka:2015bsf},
we observe that as the gauge coupling becomes stronger,
the magnetic flux of the degree $Q$ baby Skyrmions grows from 0 to $2\pi Q/g$, i.e. in the strong coupling regime
the magnetic flux is quantized, see the left bottom plot in Fig.~\ref{overg}.

We note here that since in the Chern-Simons model the electric charge of the configuration is proportional to the
magnetic flux, $q=c \Phi$, effective quantization of one quantity means that the second quantity also becomes quantized.
This effect is not topological, unlike the soliton charge $Q$,
both the magnetic flux and the electric charge are not topological numbers.

For the weakly bounded potential \re{Salmi} the  binding energy is always positive, i.e. $E_1>E_{i>1}/Q$,
where $E_1$ is the energy of the one-soliton configuration at the corresponding value of the gauge coupling $g$.
Hence the energy of a given multisoliton configuration
is less than the energy of the separated single solitons and the configuration remains stable.

In the right upper plot of Fig.~\ref{overg}, we represent the dependency of the total electromagnetic energy per unit topological
charge on the gauge coupling. Similar to the case of the gauged baby Skyrme model without the Chern-Simons term
\cite{Samoilenka:2015bsf}, the electromagnetic energy increases
till $g\simeq1$ and then tends to decrease as $g$ continue to grow. As we mentioned in \cite{Samoilenka:2015bsf},
the physical reasons of that effect could be better understood if we note that the
usual rescaling of the potential $A_\mu \to g A_\mu $
leads to $F_{\mu\nu}^2 \to \frac{1}{g^2}F_{\mu\nu}^2$. Thus, in the strong coupling regime the Maxwell term is effectively removed
from the Lagrangian \re{lag}.

The pattern of evolution of the angular moment of the  lightly bound configuration is similar to the
behavior of the electromagnetic energy, it has maximum at $g\simeq0.7$ and we can assume that $J$ tends to $0$
as $g\rightarrow\infty$ since increasing of the gauge coupling
makes the solitons width and the magnetic fluxes increasingly localized.

\begin{figure}[h]
            \begin{center}
                \includegraphics[height=5cm]{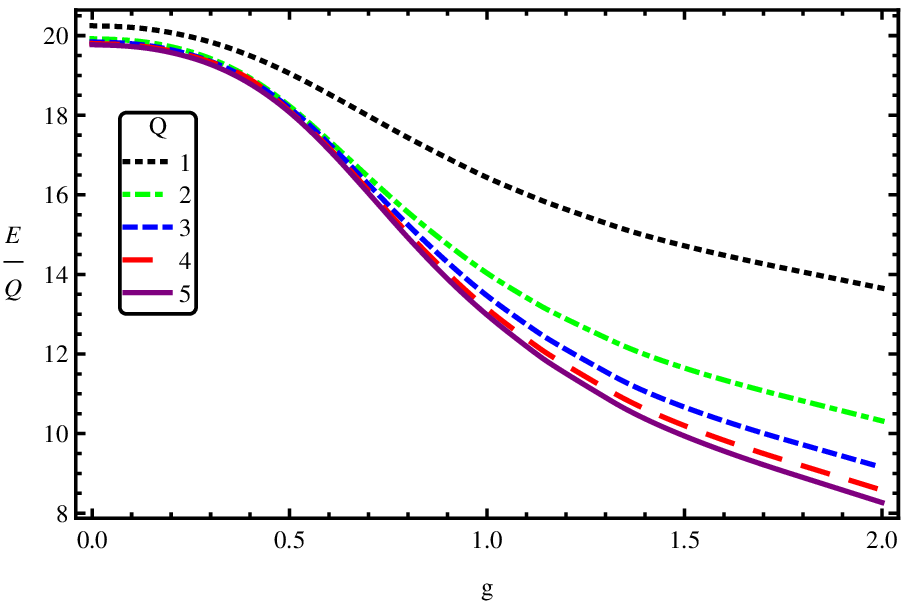}
                \includegraphics[height=5cm]{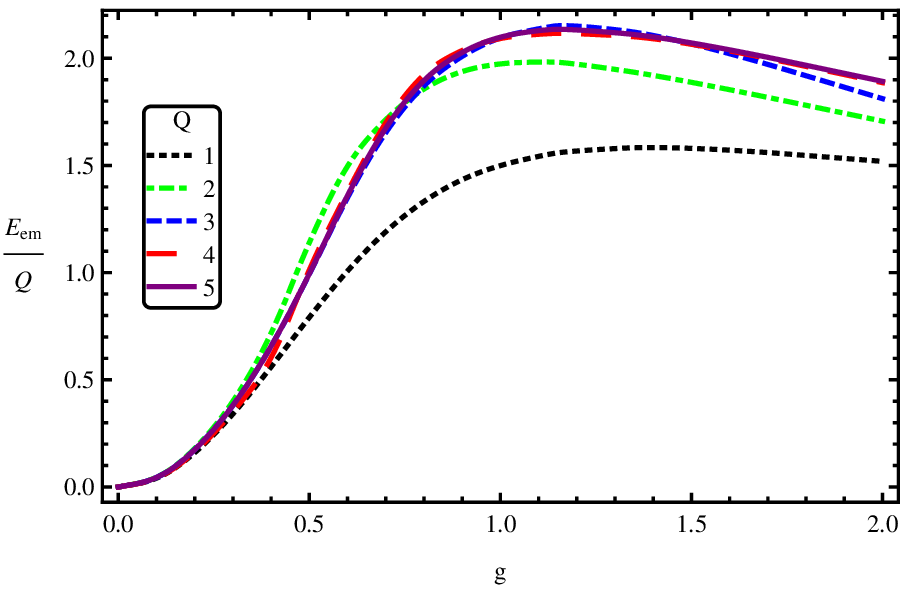}
                \includegraphics[height=5cm]{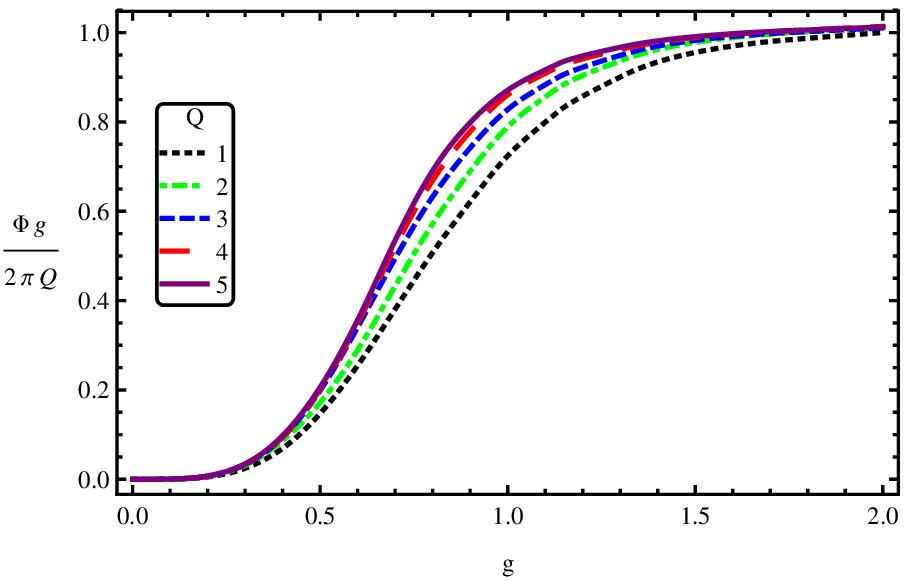}
                \includegraphics[height=5cm]{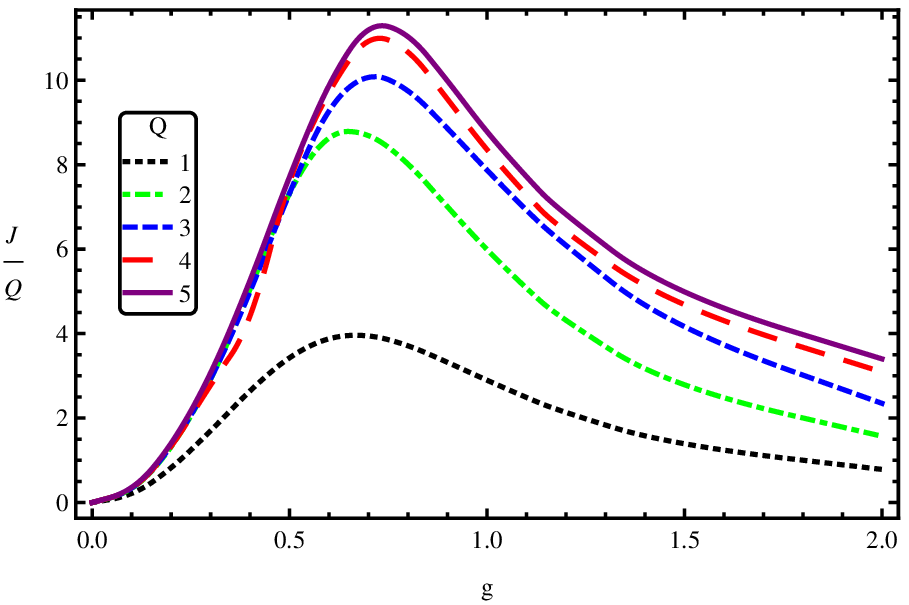}
            \end{center}
            \lbfig{overg}
            \caption{Total energy, electromagnetic energy, magnetic flux and angular momentum per charge vs gauge coupling $g$
            for the gauged planar Skyrmions with topological charges $Q=1-5$ in the model with potential \re{Salmi}
             at $\mu^2=0.1$, $\lambda=0.5$ and $\omega=0$.}
\end{figure}

In fact, the effect of the Chern-Simons coupling is to induce an electric charge of the solitons. It produces additional repulsion
between the partons. However, our numerical simulations indicate that the rotational invariance of the
multisoliton configurations is restored in the strong coupling regime. Indeed, we observe that
as the gauge coupling becomes stronger and both the electric charge and the magnetic field become quantized,
the multisoliton solution of the model \re{lag} with potential \re{Salmi} tend to form systems of
$Q=2$ rotationally invariant solitons, in agreement with the binary species model \cite{Salmi:2014hsa}.
This pattern can be seen from Fig.~\ref{table_aloof} where
we displayed the energy, the electric and magnetic field and the angular momentum
density contour plots for the $Q=2-5$ solitons at
$\omega =0$ and $g=1.5$. Further increase of the gauge coupling yields a complete
restoration of the rotational symmetry of the mutisoliton solutions. This pattern is similar to what
we observed in the absence of the electric field \cite{Samoilenka:2015bsf}, yet at some lower
values of the gauge coupling.
Thus, increase of the Chern-Simons coupling constant $c$ induces electric repulsion between the constituents,
it can be compensated by the corresponding increase of the coupling constant $g$, see  Eq. \re{eqs}.

\begin{figure}
    \begin{center}
        \begin{tabular}{|c|c|c|c|c|}
            \hline
                $Q$ & $E$ & $B$ & $|\vec E|$ & $J$ \\
            \hline
                2 & \includegraphics[height=3cm]{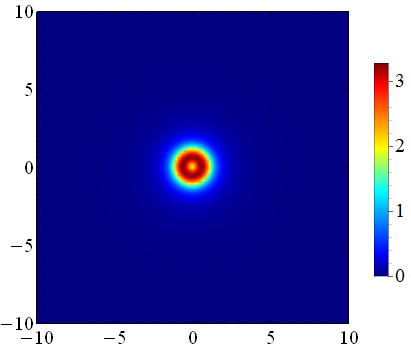} & \includegraphics[height=3cm]{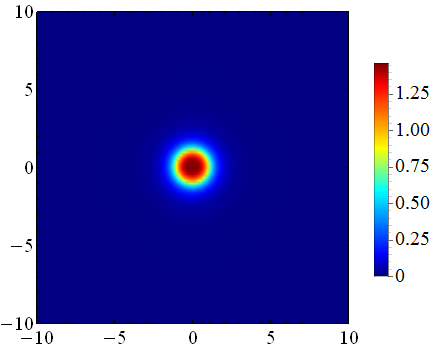} & \includegraphics[height=3cm]{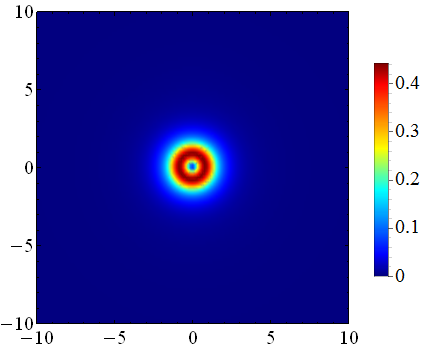} & \includegraphics[height=3cm]{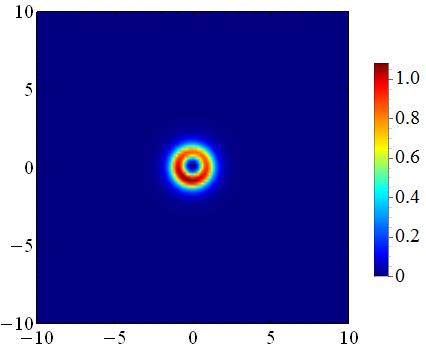} \\
            \hline
                3 & \includegraphics[height=3cm]{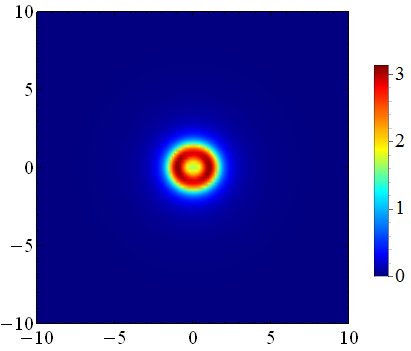} & \includegraphics[height=3cm]{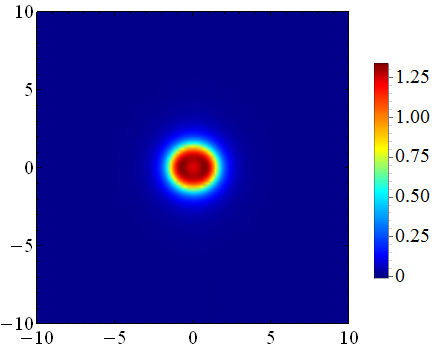} & \includegraphics[height=3cm]{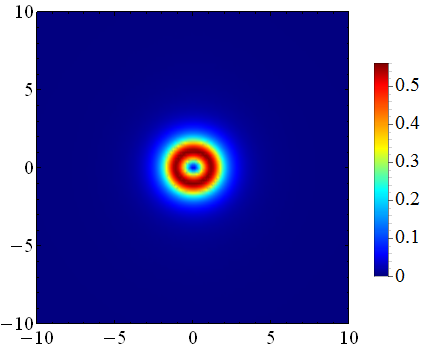} & \includegraphics[height=3cm]{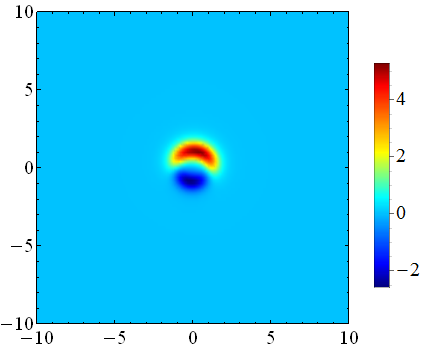} \\
            \hline
                4 & \includegraphics[height=3cm]{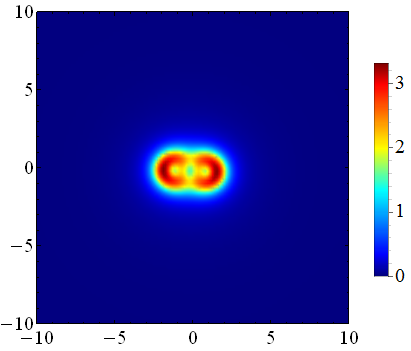} & \includegraphics[height=3cm]{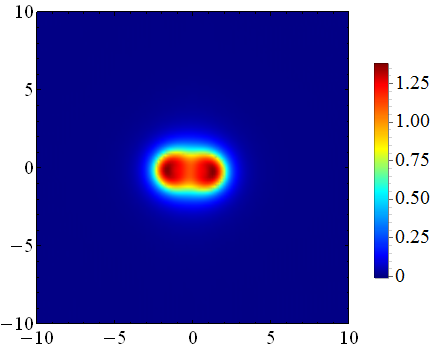} & \includegraphics[height=3cm]{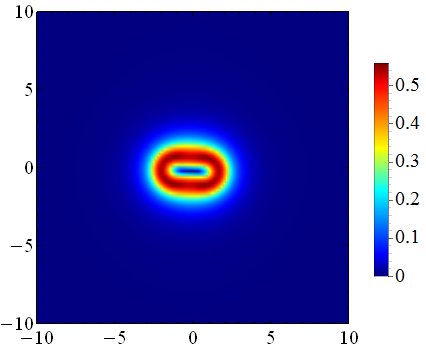} & \includegraphics[height=3cm]{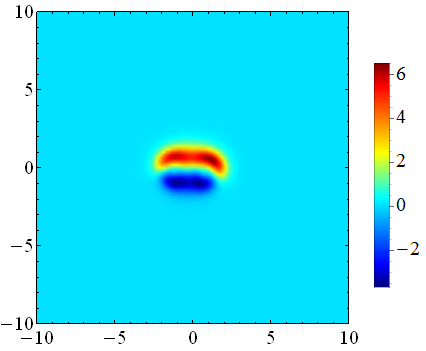} \\
            \hline
                5 & \includegraphics[height=3cm]{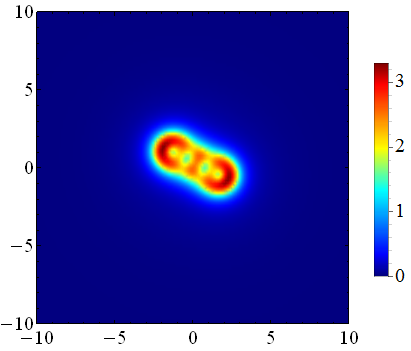} & \includegraphics[height=3cm]{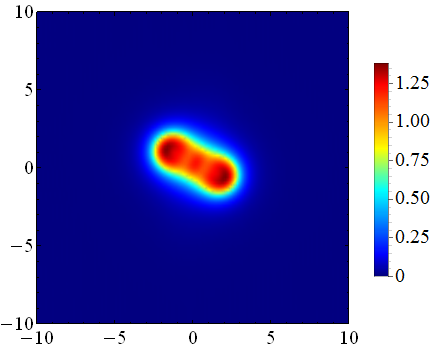} & \includegraphics[height=3cm]{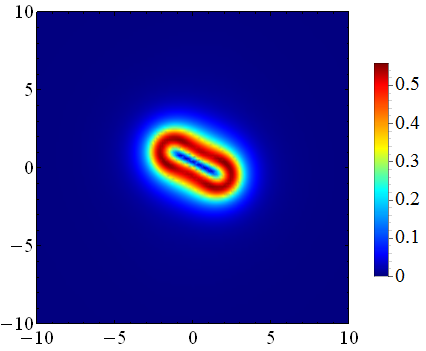} & \includegraphics[height=3cm]{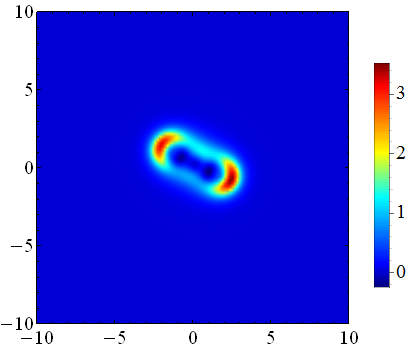} \\
            \hline
        \end{tabular}
    \end{center}
    \lbfig{table_aloof}
    \caption{Energy density, magnetic field, module of electric field and angular
    momentum density of  solutions of degrees $Q=2-5$ in the model \re{lag} with
    potential  \re{Salmi} at $\mu^2=0.1$, $\lambda=0.5$ and $\omega=0$  at $g=1.5$.}
\end{figure}

A better understanding of the complicated pattern of electromagnetic interactions in the system of solitons
requires consideration of the second of the equations \re{eqs}, which yields two Maxwell equations
\be
\begin{split}
\label{Maxwell}
&\nabla\times\vec B + c \vec E \times \vec \phi_\infty =\frac{4\pi}{c} \vec \jmath\, ; \\
&\nabla \cdot\vec E+c \vec B\cdot\vec \phi_\infty=4\pi \rho
\end{split}
\ee
where the components of the electric current
$\vec\jmath=(\jmath_1,\jmath_2,0)$ are:
\be
\begin{split}
\label{current}
    \jmath_1&=-g^2 A_1 \left[1-(\phi_3)^2+(\pa_i \phi_3)^2\right] + g \left\{(\vec\phi\times\pa_1\vec\phi)_3+
    \pa_2\phi_3(\pa_2\vec\phi\cdot\vec\phi\times\pa_1\vec\phi)\right\}\, ;\\
\jmath_2&=g^2 A_1 \pa_1\phi_3 \pa_2\phi_3 + g \left\{(\vec\phi\times\pa_2\vec\phi)_3+
\pa_1\phi_3(\pa_1\vec\phi\cdot\vec\phi\times\pa_2\vec\phi)\right\}
\end{split}
\ee
and the electric charge density is $\rho=-\frac{g^2}{4\pi}(\omega+A_0)\left[1-(\phi_3)^2+(\pa_i \phi_3)^2\right]$.
The second equation in \re{Maxwell} is just another form of the Gauss law.

Clearly, inversion of the sign of $\omega$ does not affect the current \re{current}, however the charge density $\rho$
is changing by $\delta \rho=\frac{g^2}{2\pi}\omega[(\phi_1)^2+(\phi^2)^2 + (\pa_i \phi_3)^2]$.

Let us now consider how the properties of the multisoliton configurations depend on the value of the parameter $\omega$,
which yields the electric potential in the system.

In the Fig.~\ref{overw}, we present the results of the
analysis of the dependence of the
characteristics of configuration on the rescaled
parameter $\omega \in [-1,1]$ . We fixed an intermediate value of the gauge coupling $g=0.3$.

First, we observe that the for all range of values of the parameter $\omega$ the multisoliton configuration in
the above-mentioned system consists of individual
partons of unit charge. As $|\omega|\rightarrow1$ the electric repulsion becomes strong and
the solitons become unstable, thus all quantities grow rapidly in magnitude and diverge.
However the binding energy  at $\omega \to -1$ is larger than at $\omega \to 1$, see
the left upper plot of Fig.~\ref{overw}.

As said above, the Chern-Simons term violates the $P$ and $T$ invariance of the system, therefore
the structure of the configurations is not symmetric with respect to reflections $\omega \to -\omega$. Indeed, it is seen in
Fig.~\ref{overw}, lower row, which displays the dependencies of the integrated magnetic flux\footnote{Recall that the magnetic
flux is related to the total electric charge of the configuration as $q=c\Phi$.}  and angular momentum on $\omega$,
both quantities are vanishing at some non-zero positive value of $\omega$. They become negative and continue to decrease monotonically as
$\omega$ grows. Notably, the electromagnetic energy of the $Q=2$ configuration per unit charge is higher that for other multisolitons,
see the right upper plot of Fig.~\ref{overw}.
\begin{figure}[h]
            \begin{center}
                \includegraphics[height=5cm]{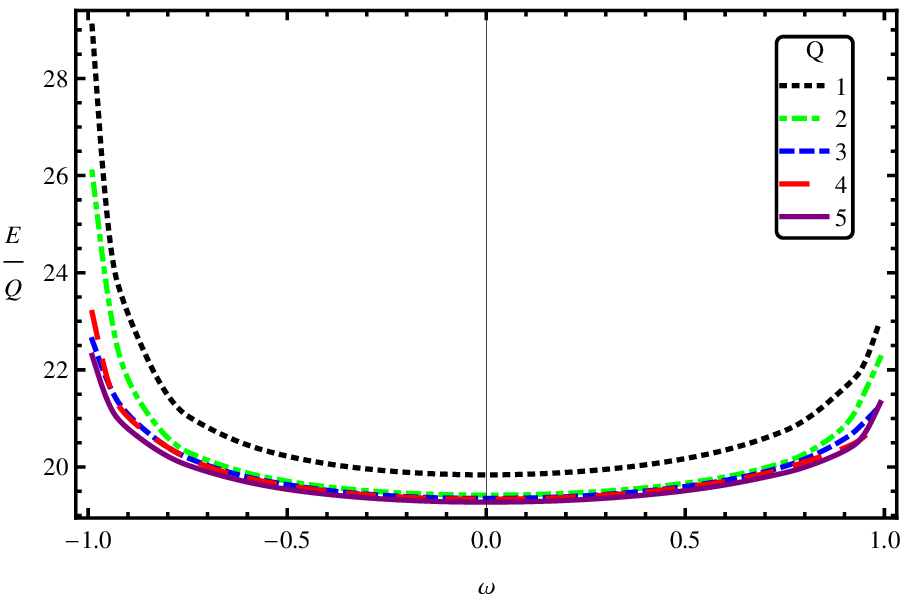}
                \includegraphics[height=5cm]{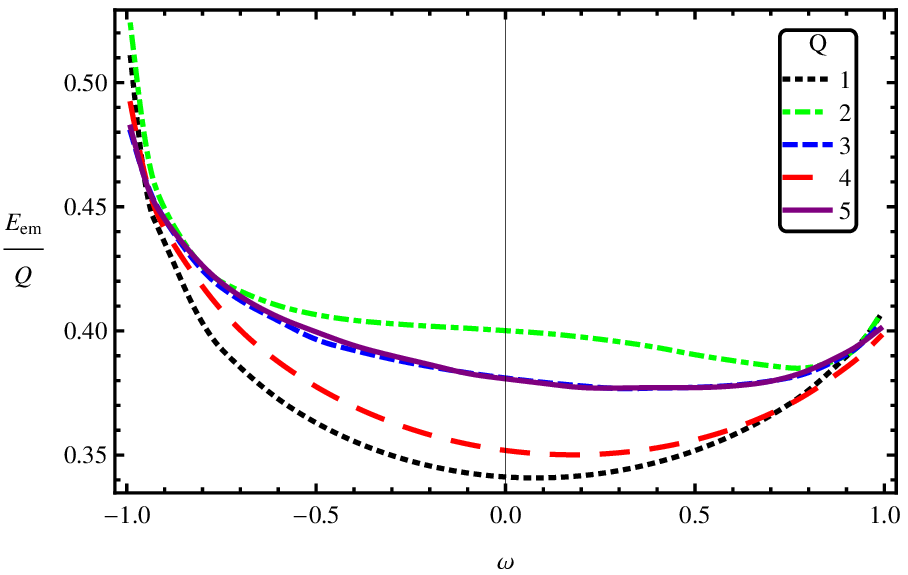}
                \includegraphics[height=5cm]{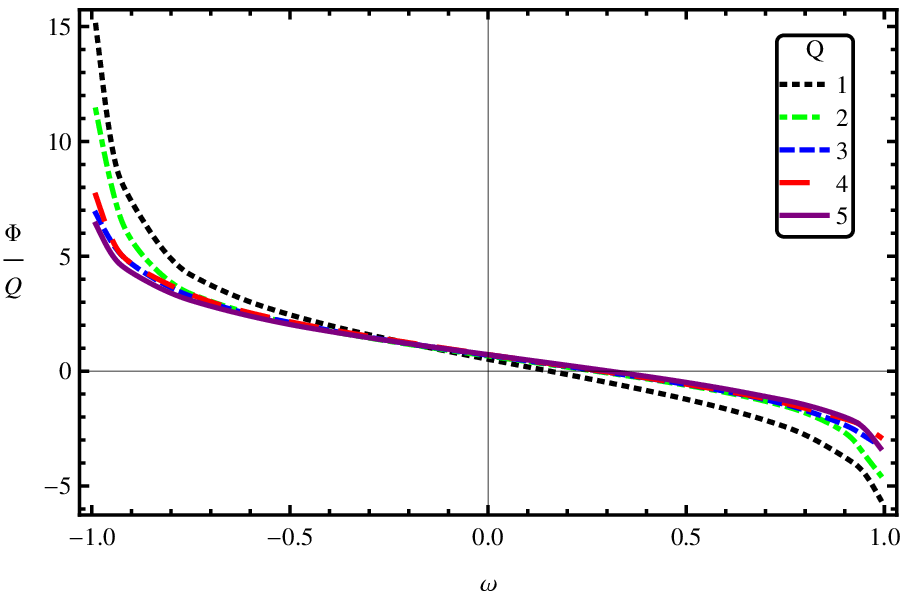}
                \includegraphics[height=5cm]{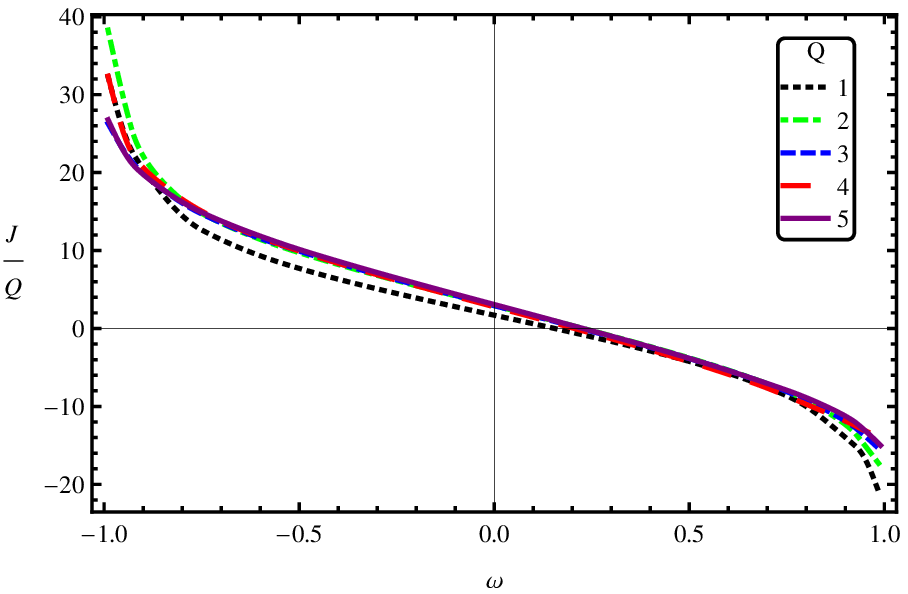}
            \end{center}
            \lbfig{overw}
            \caption{
Total energy, electromagnetic energy, magnetic flux and angular momentum per charge vs $\omega$
            for the gauged planar Skyrmions with topological charges $Q=1-5$ in the model with potential \re{Salmi}
            at $\mu^2=0.1$, $\lambda=0.5$ and $g=0.3$.}
\end{figure}

In Fig.~\ref{aloof-2} we displayed the energy, the electric and magnetic field  density contour plots for the $Q=2$ solitons at
$\omega =0.99$ and $\omega=-0.99$. Clearly, in the latter case the separation between the constituents is a bit smaller.
Peculiar feature of the magnetic field distribution is that for $\omega=0.99$ the positive magnetic flux exhibits a peak
at the center of the Skyrmion. However  this peak
is screened by the circular wall of negative magnetic flux, so the total flux is negative. For $\omega=-0.99$ the magnetic flux
has a single positive maximum at the center of the soliton.

\begin{figure}[h]
           \begin{center}
                \includegraphics[height=4.cm]{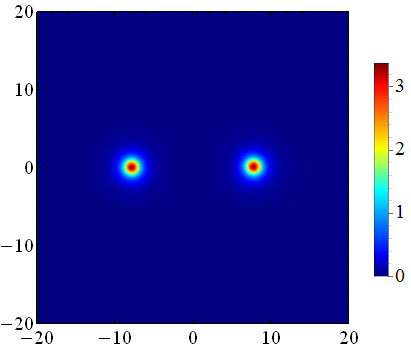}
                \includegraphics[height=4.cm]{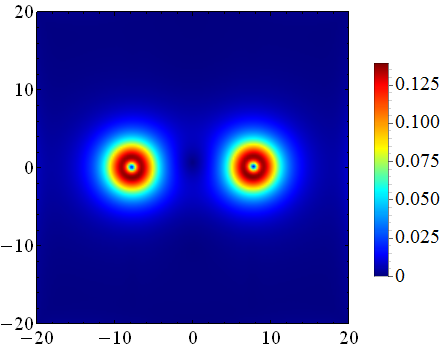}
                \includegraphics[height=4.cm]{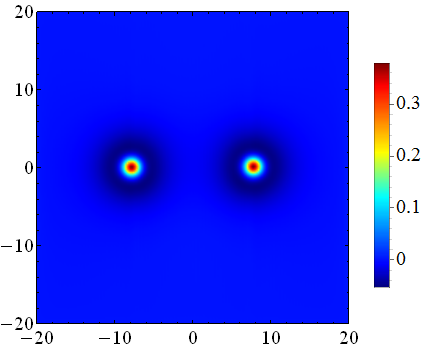}
                \includegraphics[height=4.cm]{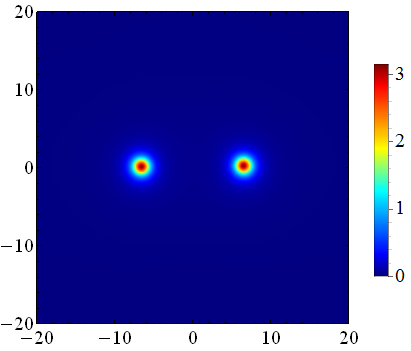}
                \includegraphics[height=4.cm]{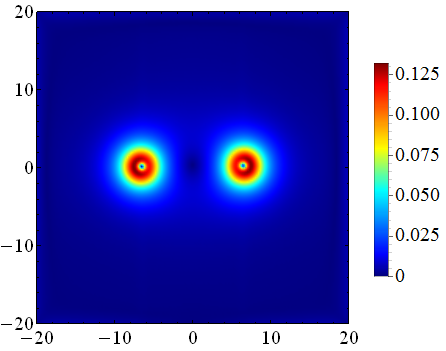}
                \includegraphics[height=4.cm]{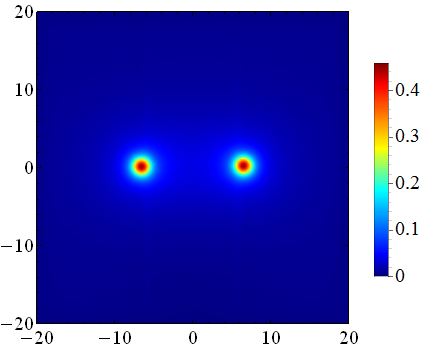}
            \end{center}
            \lbfig{aloof-2}
\caption{Contour plots of the energy density (left column) electric field  (middle column) and the magnetic field
distributions (right column)  for $Q=2$ gauged planar Skyrmions in the model \re{lag} with potential \re{Salmi} at
$\mu^2=0.1$, $\lambda=0.5$, $g=0.3$ and $\omega=0.99$ (upper row), $\omega=-0.99$ (lower row).}
\end{figure}

\subsection{Double vacuum potential}

Next we consider the model \re{lag} with the double vacuum potential \re{double}. In this case
the bound \re{restr-pot} becomes
\be
    |\omega|<\frac{1}{g}\sqrt{-\frac{\pa V}{\pa \phi_3}\bigg|_{\phi_3\rightarrow1}}=\frac{\sqrt{2\mu^2}}{g}
\ee
Thus, for comparative consistency of our consideration, we fix $c=1$, $\mu^2=0.1$ and rescale the parameter $\omega$ as
$\omega\rightarrow\frac{\sqrt{2\mu^2}}{g}\omega$, so that $\omega\epsilon[-1,1]$. Again, we
consider multisoliton configurations of degrees $Q=1-5$.

As  $\omega=0$ the solutions possess the rotational invariance as in the absence of the Chern-Simons term.
However as $\omega$ start to vary, the situation changes.
While the unit charge Skyrmion remains rotationally invariant for all range of values of $\omega$, the
repulsive electromagnetic interaction may
affect the structure of multisolitons at large negative values of this parameter.

In Fig.~\ref{table_double_en}  we exhibited the contour energy density plots of the
gauged baby Skyrmions in the model \re{lag} with double vacuum potential for $1\le Q\le 5$
at some set of values of $\omega$. Indeed, electromagnetic repulsion becomes stronger as $\omega \to -1$ and,
as electric charge of the solitons  becomes large enough, the size of the configuration starts to increase and
the  multi-soliton configuration features broken rotational invariance.
Then the global minima are symmetric with respect to the dihedral group $D_Q$, see the first
column in Fig.~\ref{table_double_en}.
For positive values of  $\omega$ we still can observe the
local minima consisting of several  constituents, see Fig.~\ref{table_double}.

%\begin{figure}[h]
%            \begin{center}
%                \includegraphics[height=6cm]{d_ch=2w=o99(1+1).png}
%                \includegraphics[height=6cm]{d_ch=3w=o99(1+2).png}
%                \includegraphics[height=6cm]{d_ch=4w=o9(2+2).png}
%                \includegraphics[height=6cm]{d_ch=5w=o9(2+3).png}
%            \end{center}
%            \lbfig{plus}
%            \caption{Energy density plot for $1+1$, $1+2$, $2+2$ and $2+3$ solitons respectively with $g=0.3$ and $\omega=0.99$ for first two of them and %$\omega=0.9$ for the last two of them}
%\end{figure}

%\begin{figure}[h]
%            \begin{center}
%                \includegraphics[height=6cm]{d_ch=2w=-o99fig.png}
%                \includegraphics[height=6cm]{d_ch=3w=-o99fig.png}
%                \includegraphics[height=6cm]{d_ch=4w=-o9.png}
%                \includegraphics[height=6cm]{d_ch=5w=-o9.png}
%            \end{center}
%            \lbfig{dihedral}
%            \caption{Energy density plot for $Q=2,3,4,5$ with $g=0.3$ and $\omega=0.99$ for first two of them and $\omega=0.9$ for the last two of them}
%\end{figure}

In Fig.~\ref{overw_double} we have plotted the dependencies of energy, electromagnetic energy, magnetic flux and angular
momenta of the static soliton solutions of the model \re{lag} with double vacuum potential \re{double}
as functions of $\omega$ at $g=0.3$.
Qualitatively, the results are rather similar to the pattern reported above for the model with weakly bounded potential,
cf. Fig.~\ref{overw}. A significant difference is related with the behavior of the solutions at the negative values of $\omega$.
Indeed, it is seen in the left upper plot of Fig.~\ref{overw_double}, at some critical negative value of $\omega$
we observe crossings in $E(\omega)$ curves. First, the energy of the  multi-Skyrmion configuration becomes higher than the energy
of the system of $Q$ charge one baby Skyrmions and the configurations are unstable with respect to decay into
constituents. For some range of values of parameters of the model another channels of decay are also possible, for example
the configuration may decay into set of $Q=2$ rotationally invariant solitons, as in the binary species model \cite{Salmi:2014hsa}.
Then, as the value of the parameter $\omega$ continues to decrease approaching lower bound $-1$, the second crossing in
$E(\omega)$ curves is observed, the system becomes bounded in a new configuration with $E_1 > E_{i>1}/Q$.

Another indication of this instability is the curve $J/Q(\omega)$ displayed in Fig.~\ref{overw_double}, right bottom plot. As we can see,
the momentum curve does not have a monotonic behavior, as $\omega$ decreases below the corresponding negative critical value and the
rotational symmetry of the configuration becomes broken, the angular momentum starts to decrease rapidly.

\begin{figure}[h]
            \begin{center}
                \includegraphics[height=5cm]{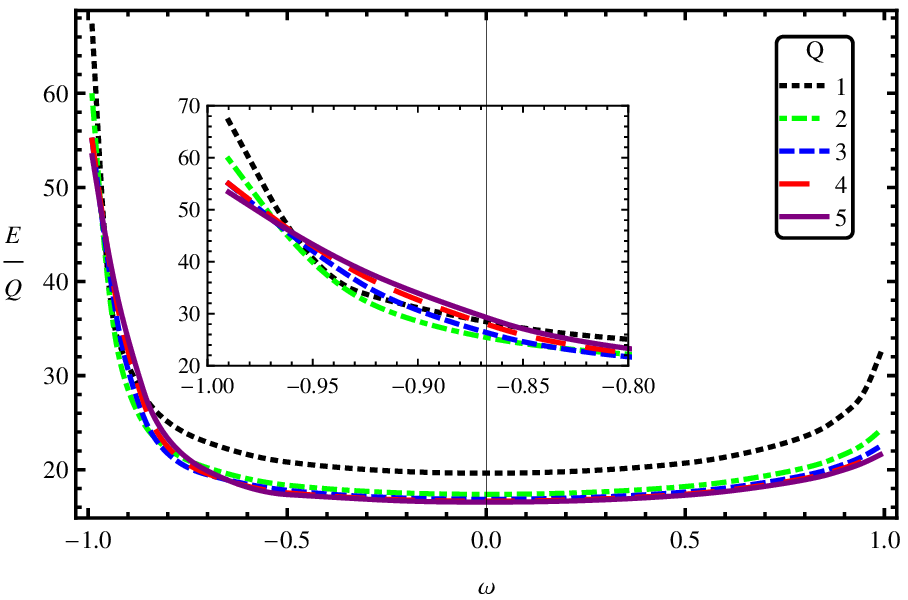}
                \includegraphics[height=5cm]{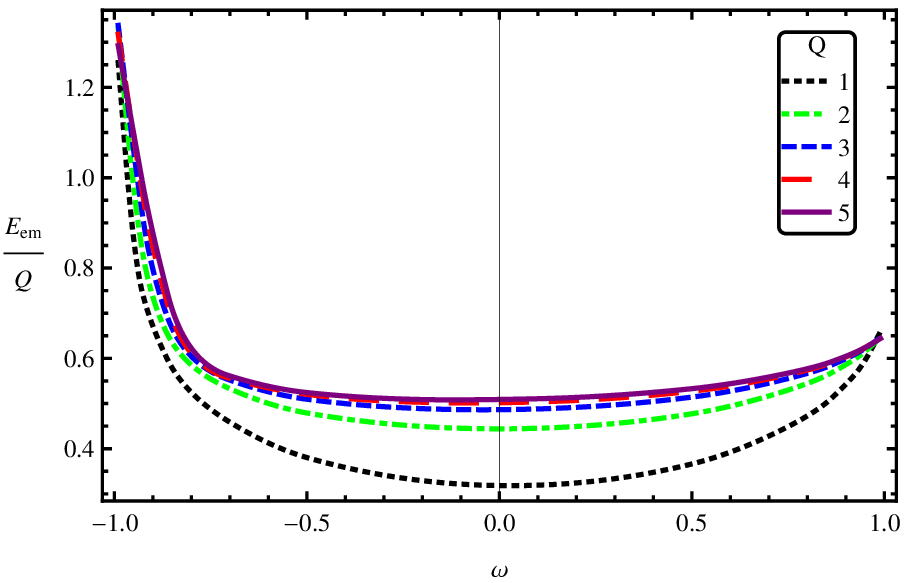}
                \includegraphics[height=5cm]{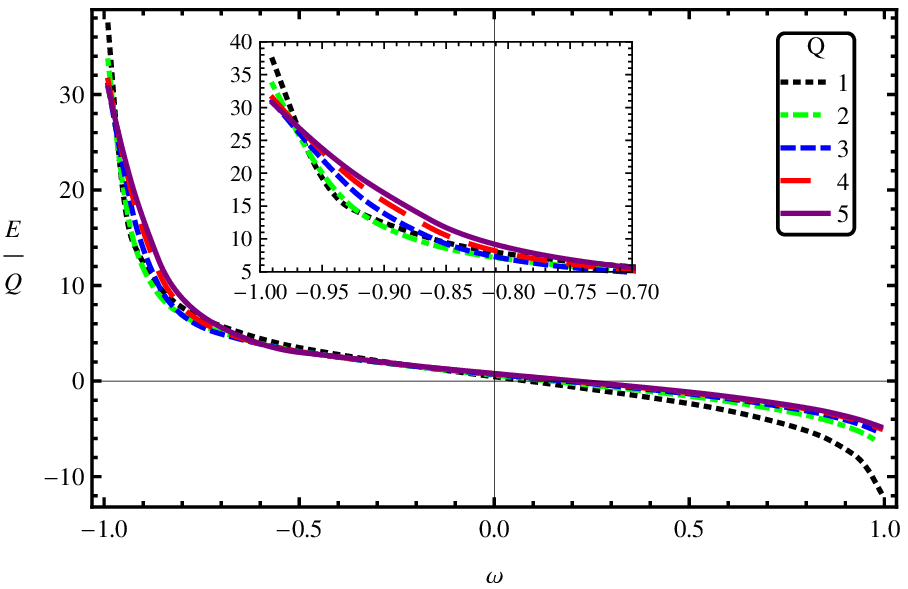}
                \includegraphics[height=5cm]{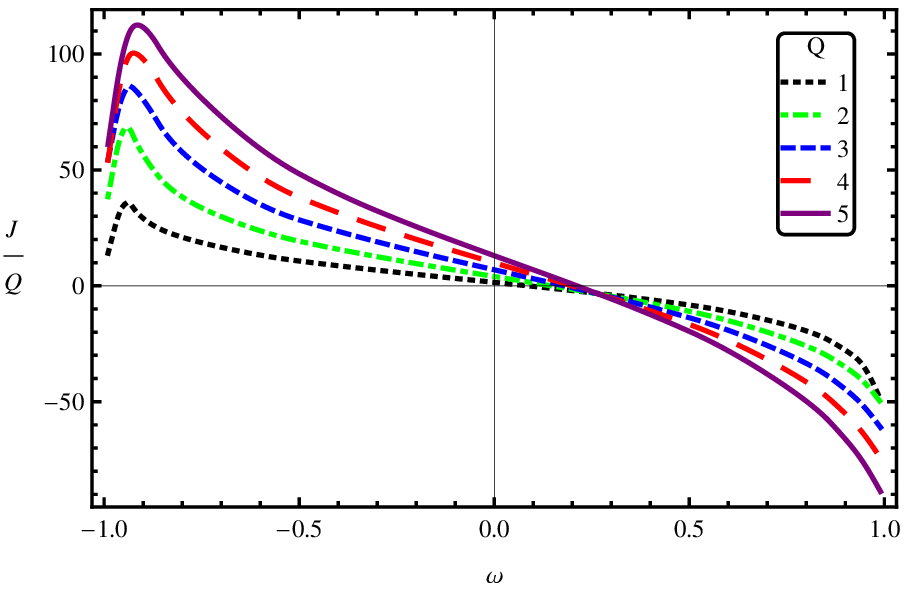}
            \end{center}
            \lbfig{overw_double}
            \caption{Total energy, electromagnetic energy, magnetic flux and angular momentum per charge for $Q=1-5$
            with double vacuum potential \re{double} and $g=0.3$.For $Q=1$ our results agree well with those presented in \cite{Loginov}.}
\end{figure}

In Fig.~\ref{table_double} we summarized our results for critical behavior of the multisolitons of the model \re{lag} with the
double vacuum potential at large positive and negative values of $\omega$. The $Q=1$ Skyrmion remains rotationally invariant, however
the size of the soliton at $\omega=-0.99$ is much larger than at $\omega=0.99$, the energy is less concentrated at the center of the soliton.
The distribution of the electric field of the soliton in both cases is featuring an annular shape. Similar to the case of the model with
weakly bounded potential \re{Salmi} that we considered above, at large negative values of $\omega \to  -1$,
the distribution of the magnetic field has a sharp peak at center of the soliton. For positive values of $\omega$ this peak is also presented,
however it becomes screened by the circular wall of negative magnetic flux, thus the total magnetic flux through the plane is negative.

In the sector of degree two we can also see the difference between the pattern of critical behavior of the
solitons in the limit of large positive and negative values of $\omega$. As $\omega \to -1$, the electromagnetic repulsion still remains not
strong enough to break the configuration into two soltions of unit charge. However the rotational invariance is broken and the energy density
distribution has a shape of an oval cup. In the opposite limit $\omega \to 1$, the electromagnetic repulsion may broke the configuration into
pair of individual solitons, whose relative orientation corresponds to the attractive scalar channel. This configuration, however corresponds
to the local minimum, the global minimum in this sector is given by the rotationally invariant hedgehog solution.
Similarly, the  distributions of all quantities for the $Q=3$ baby Skyrmion at $\omega \to -1$ correspond to the system of
three solitons of unit charge placed at the vertices of an equilateral triangle. As $\omega\to 1$,
the rotational invariance of the global minimum in this sector is not violated, although due to electromagnetic repulsion,
the size of the configuration is much larger than at $\omega=0$.  The $1+2$ configuration remains a local minimum in this limit.
Similar scenario holds in the sector of degree five.

\begin{figure}[h]
            \begin{center}
                \includegraphics[height=6cm]{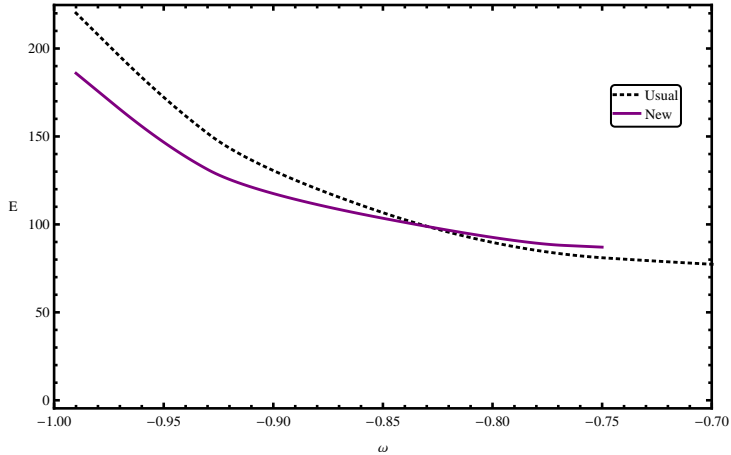}
            \end{center}
            \lbfig{flower_usual}
            \caption{Energy of Q=4 solutions vs $\omega$ in the model \re{lag}  with double vacuum
            potential \re{double} at $g=0.3$ for old (dotted) and new (solid line) configurations.}
\end{figure}

In the sector of degree four in the limit $\omega \to -1$ we found new rather peculiar solution, whose magnetic properties are different from
the pattern we described above. Indeed, there is a configuration with dihedral $D_4$ symmetry of all characterictics, including
the magnetic flux distribution. This is a bounded system of four $Q=1$ Skyrmions, placed at the vertices of a square.
However, for a given set of parameters of the model, this solution corresponds to the local minimum
as  $\omega\lesssim-0.85$. Another solution of a new type appears as a global minimum instead, its magnetic field distribution is different,
there is a sharp negative minimum at the center, surrounded by four segments of positive values of the magnetic field. The energy density
is featuring the discrete dihedral symmetry with a sharp peak at the center of the configuration. This configuration could only
be found in the ranges $\omega\in[-0.75,-1]$. At $\omega = -0.85$ both configurations of degree four are degenerate in energy,
see  Fig.\ref{flower_usual}.
Beyond this point the old solution continue to exist as a saddle point of the total energy functional\footnote{Note that this type of bifurcations
was observed in the system of isospinning solitons of degree four in the Faddeev-Skyrme model \cite{Harland:2013uk}.}.

As $\omega \to 1$, we found two different $Q=4$ solutions, one, which represent a global minimum, is rotationally invariantly hedgehog,
another configuration, which energy is slightly higher, represent a system of well separated charge 2 solitons in the attractive scalar
channel.

\begin{figure}
    \begin{center}
        \begin{tabular}{|c|c|c|c|c|c|}
            \hline
                Q & \multicolumn{5}{c|}{$\omega$}  \\
            \hline
                1 & -0.99 & -0.4 & 0 & 0.4 & 0.99\\
                & \includegraphics[width=3cm]{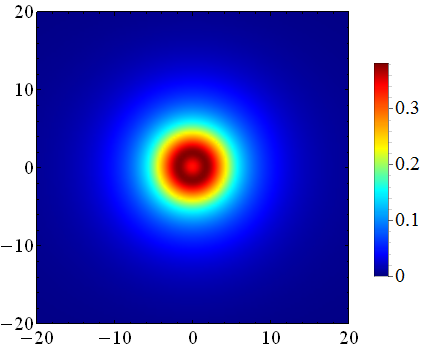} & \includegraphics[width=3cm]{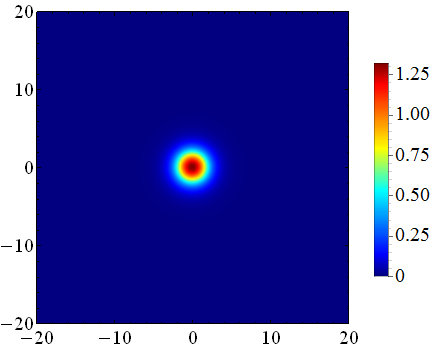} & \includegraphics[width=3cm]{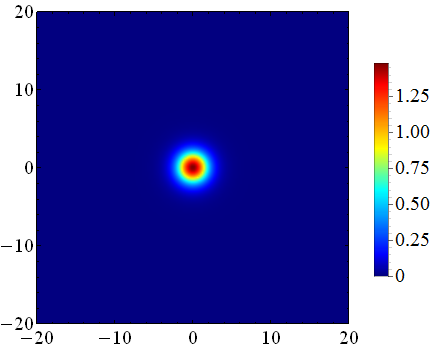} & \includegraphics[width=3cm]{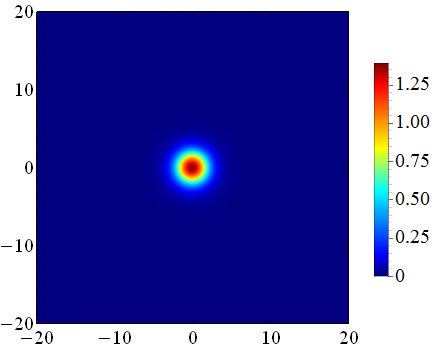} & \includegraphics[width=3cm]{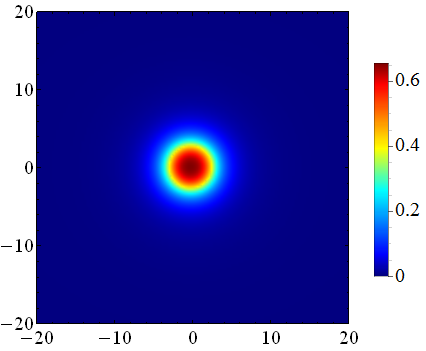} \\
            \hline
                2 & -0.99 & -0.4 & 0 & 0.4 & 0.99\\
                & \includegraphics[width=3cm]{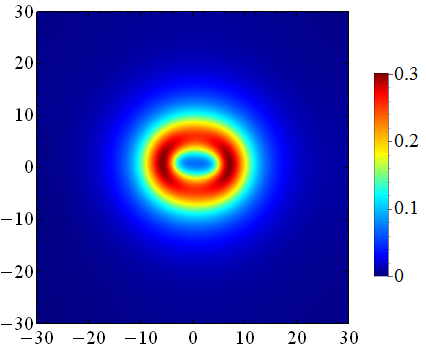} & \includegraphics[width=3cm]{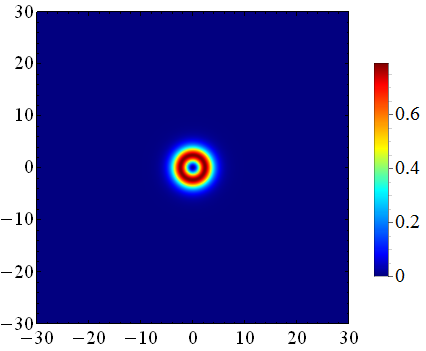} & \includegraphics[width=3cm]{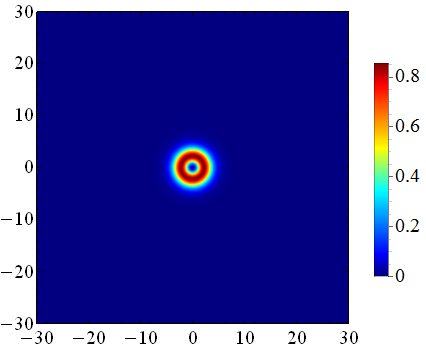} & \includegraphics[width=3cm]{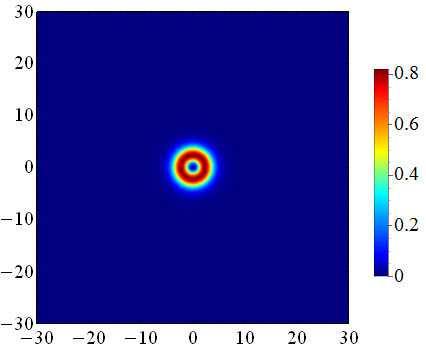} & \includegraphics[width=3cm]{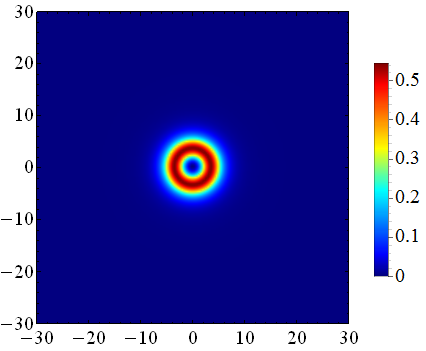} \\
            \hline
                3 & -0.99 & -0.4 & 0 & 0.4 & 0.99\\
                & \includegraphics[width=3cm]{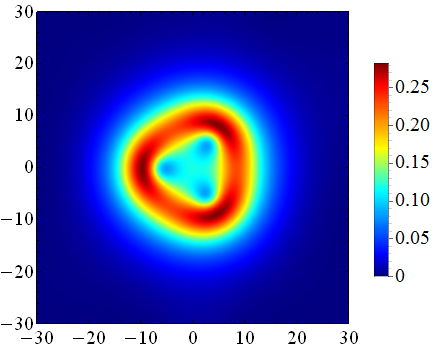} & \includegraphics[width=3cm]{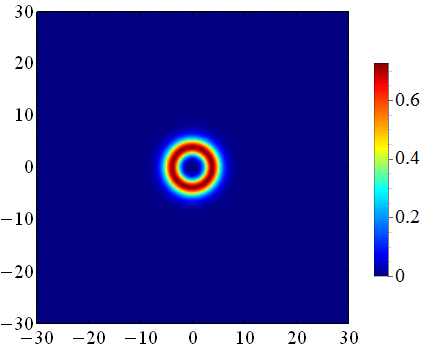} & \includegraphics[width=3cm]{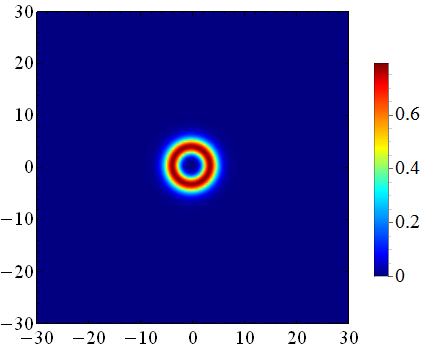} & \includegraphics[width=3cm]{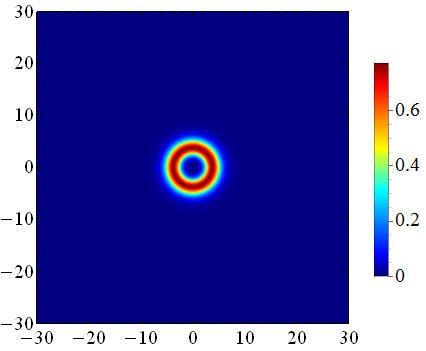} & \includegraphics[width=3cm]{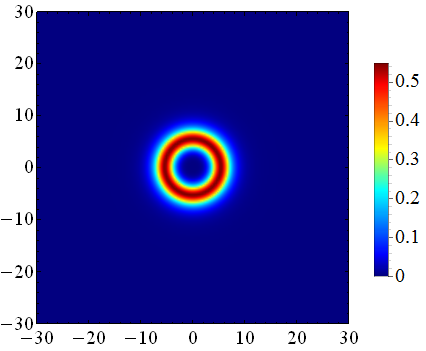} \\
            \hline
                4 & -0.99 & -0.4 & 0 & 0.4 & 0.99\\
                & \includegraphics[width=3cm]{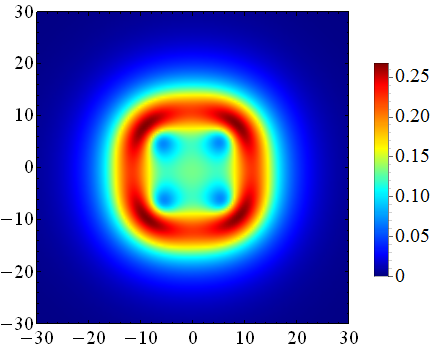} & \includegraphics[width=3cm]{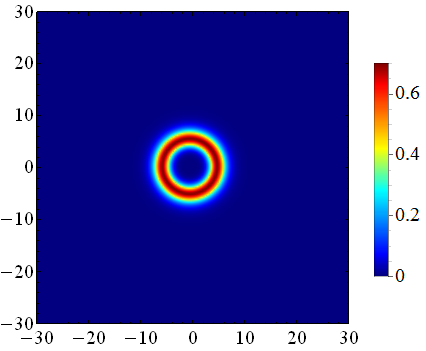} & \includegraphics[width=3cm]{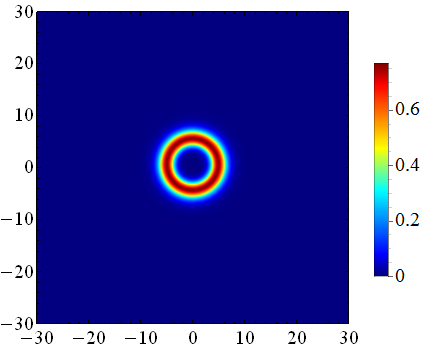} & \includegraphics[width=3cm]{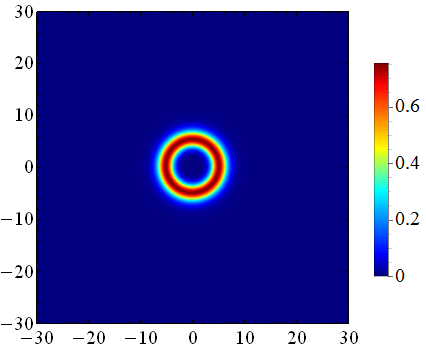} & \includegraphics[width=3cm]{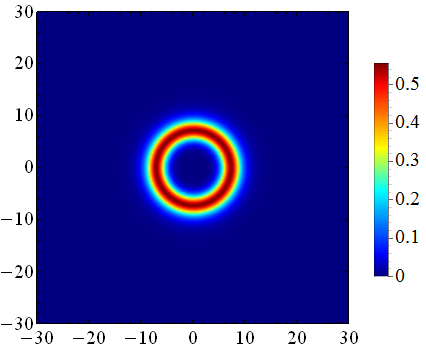} \\
            \hline
                5 & -0.99 & -0.4 & 0 & 0.4 & 0.99\\
                & \includegraphics[width=3cm]{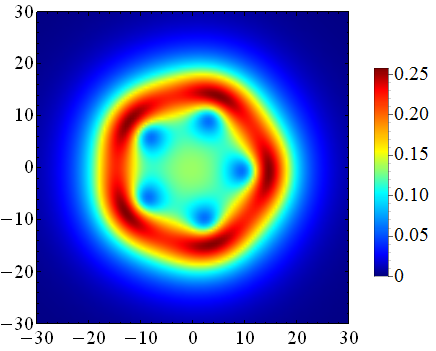} & \includegraphics[width=3cm]{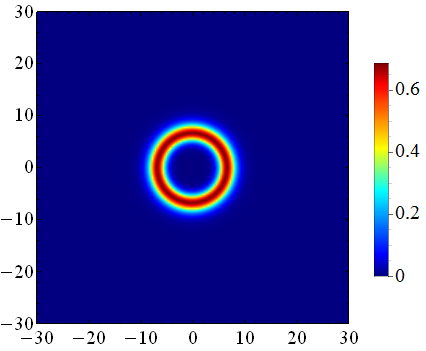} & \includegraphics[width=3cm]{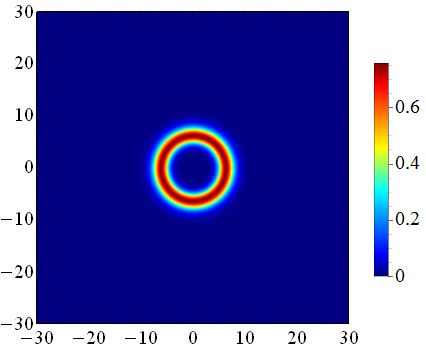} & \includegraphics[width=3cm]{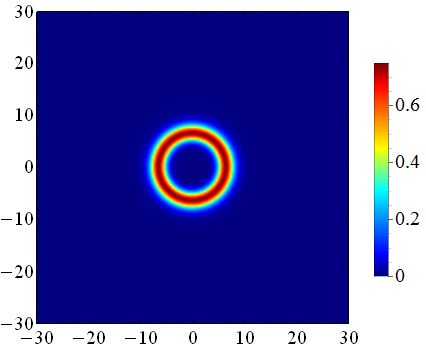} & \includegraphics[width=3cm]{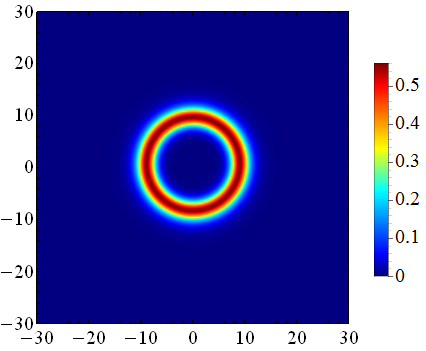} \\
            \hline
        \end{tabular}
    \end{center}
    \lbfig{table_double_en}
    \caption{Energy density of solutions in model with double vacuum potential \re{double}
    for charges $Q=1-5$, $g=0.3$ and $\omega\in[-1,1]$}
\end{figure}

\begin{figure}
    \begin{center}
        \begin{tabular}{|c|c|c|c|c|c|}
            \hline
                $Q$ & $\omega$ & $E$ & $B$ & $|\vec E|$ & $J$  \\
            \hline
                1 & -0.99 & \includegraphics[height=3cm]{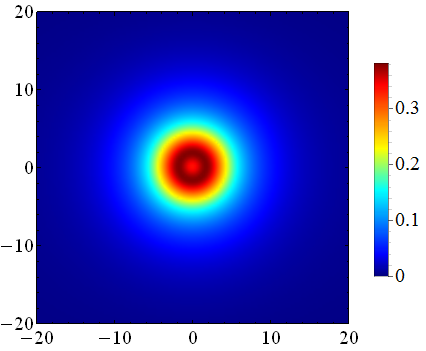} & \includegraphics[height=3cm]{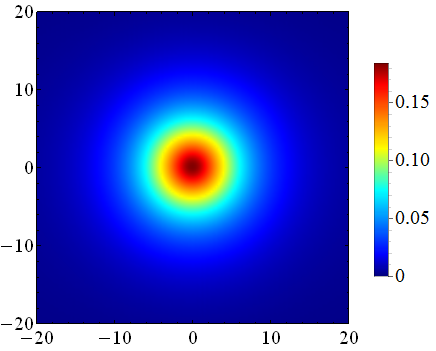} & \includegraphics[height=3cm]{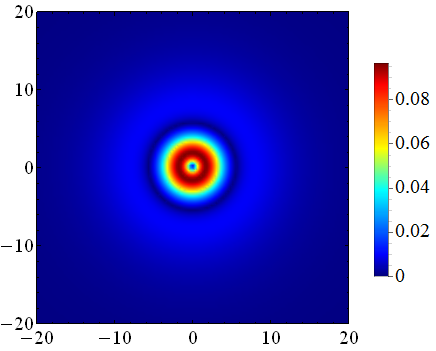} & \includegraphics[height=3cm]{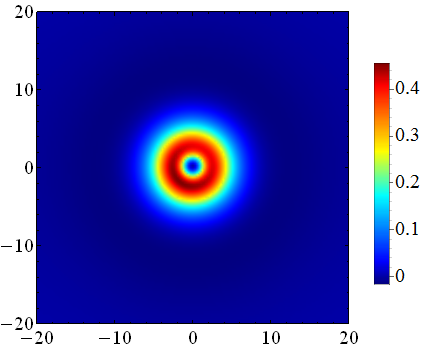} \\
            \hline
                1 & 0.99 & \includegraphics[height=3cm]{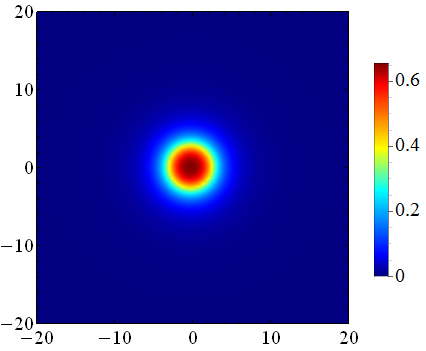} & \includegraphics[height=3cm]{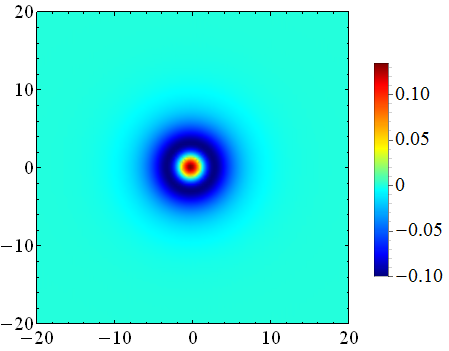} & \includegraphics[height=3cm]{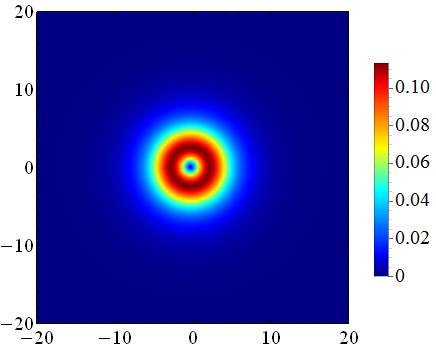} & \includegraphics[height=3cm]{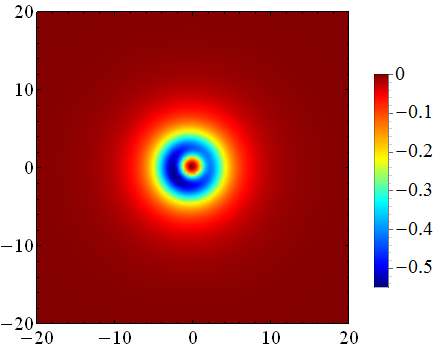} \\
            \hline
                2 & -0.99 & \includegraphics[height=3cm]{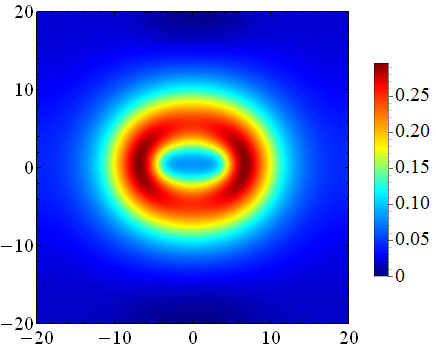} & \includegraphics[height=3cm]{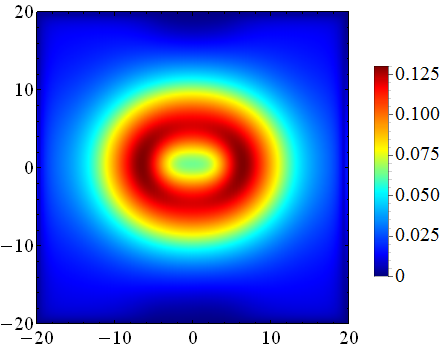} & \includegraphics[height=3cm]{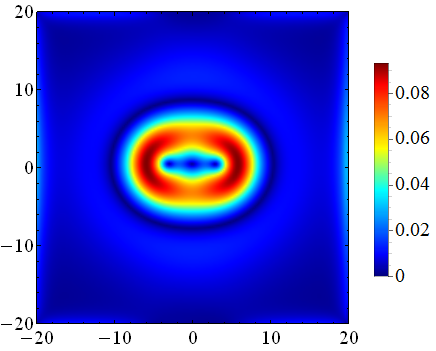} & \includegraphics[height=3cm]{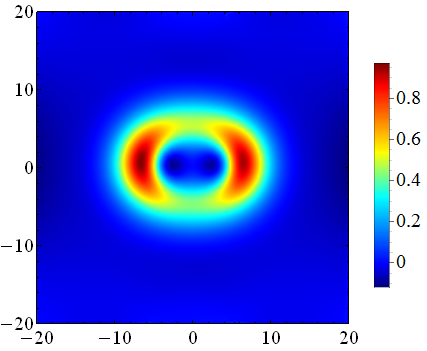} \\
            \hline
                2* & 0.99 & \includegraphics[height=3cm]{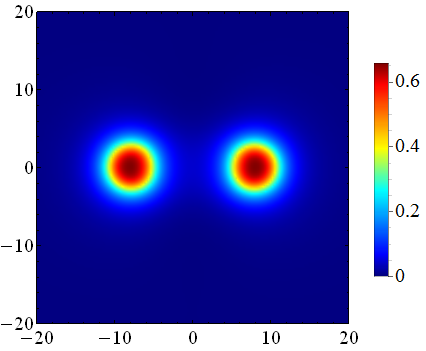} & \includegraphics[height=3cm]{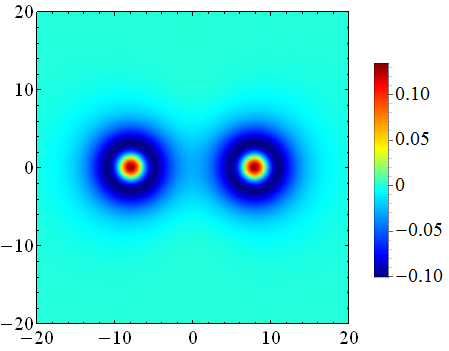} & \includegraphics[height=3cm]{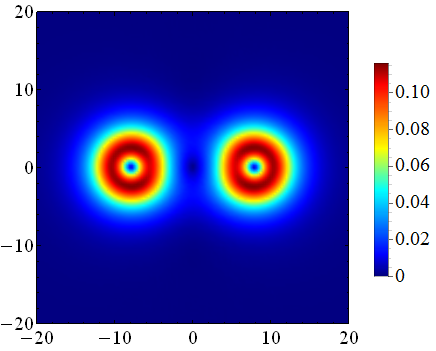} & \includegraphics[height=3cm]{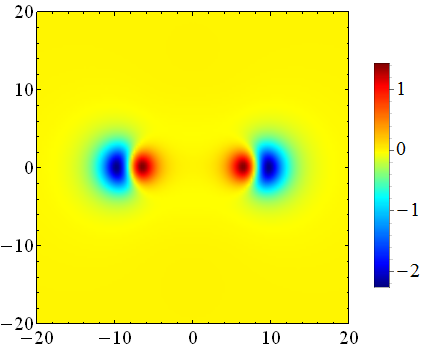} \\
            \hline
                2 & 0.99 & \includegraphics[height=3cm]{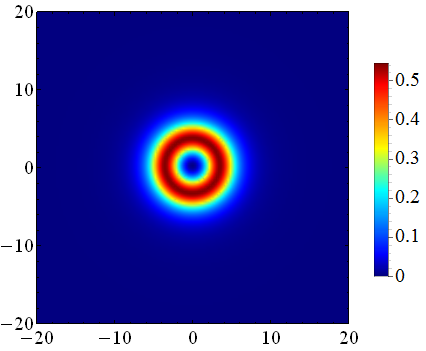} & \includegraphics[height=3cm]{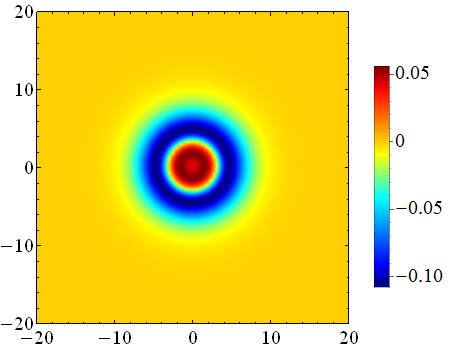} & \includegraphics[height=3cm]{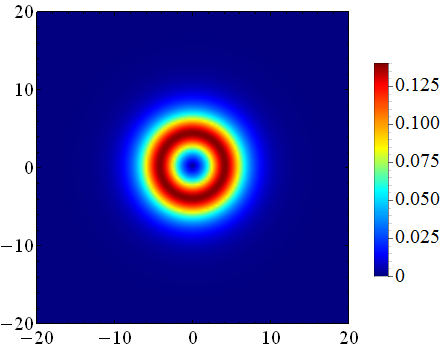} & \includegraphics[height=3cm]{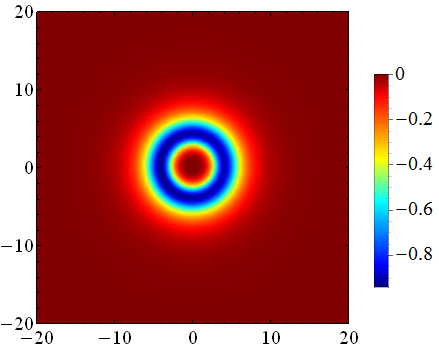} \\
            \hline
                3 & -0.99 & \includegraphics[height=3cm]{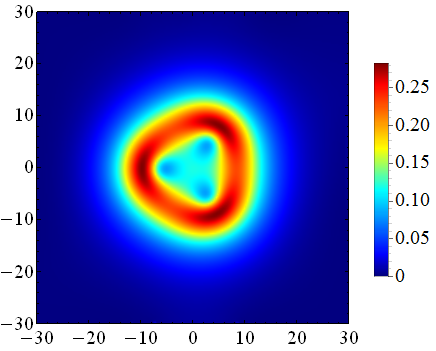} & \includegraphics[height=3cm]{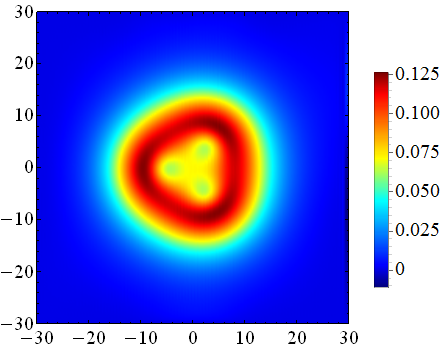} & \includegraphics[height=3cm]{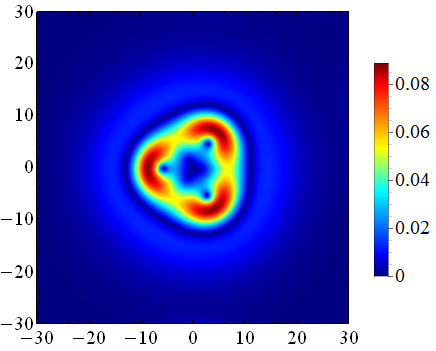} & \includegraphics[height=3cm]{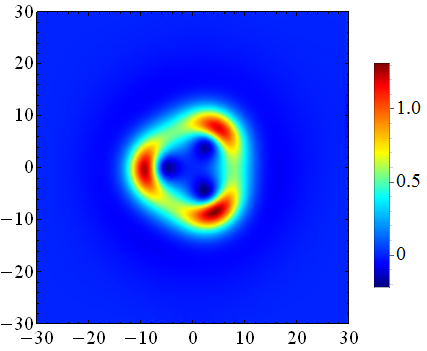} \\
            \hline
                3* & 0.99 & \includegraphics[height=3cm]{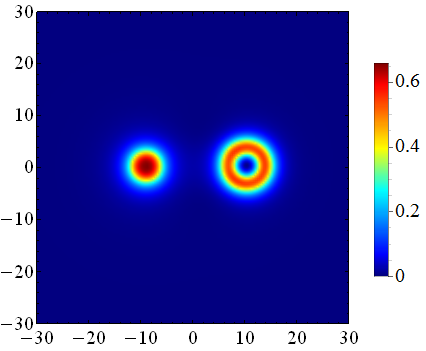} & \includegraphics[height=3cm]{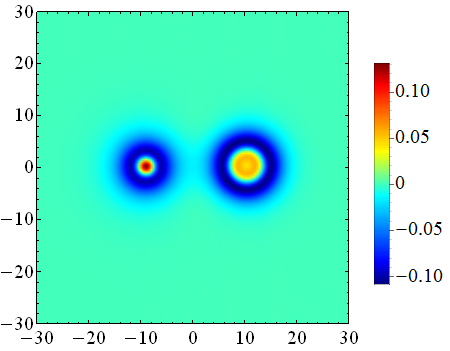} & \includegraphics[height=3cm]{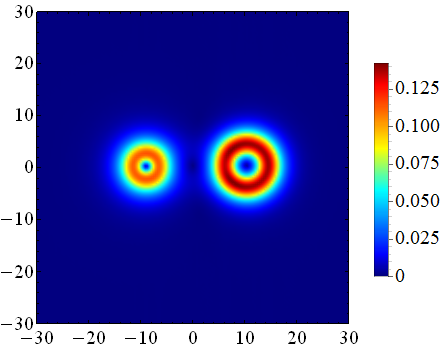} & \includegraphics[height=3cm]{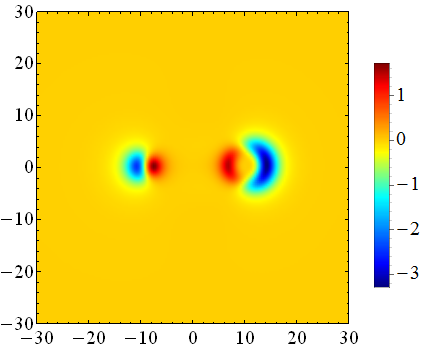} \\
            \hline
                3 & 0.99 & \includegraphics[height=3cm]{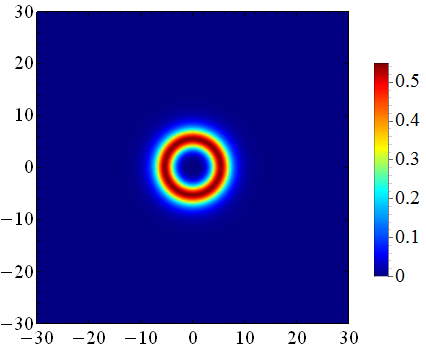} & \includegraphics[height=3cm]{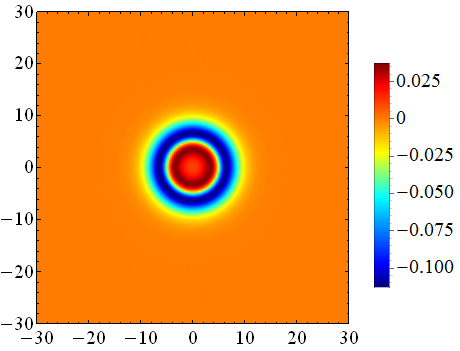} & \includegraphics[height=3cm]{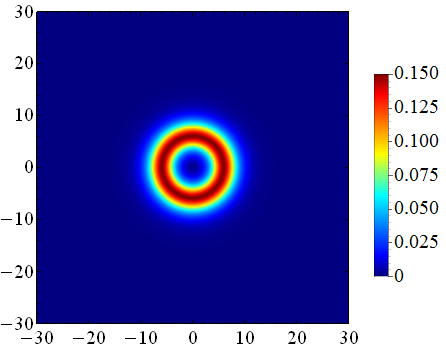} & \includegraphics[height=3cm]{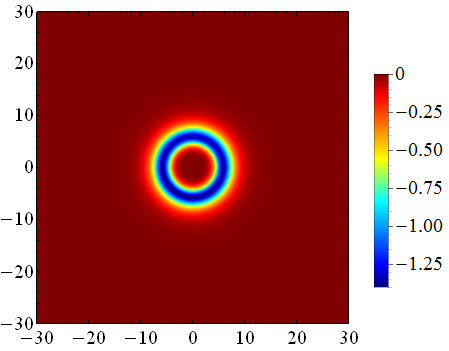} \\
            \hline
        \end{tabular}
    \end{center}
\end{figure}

\begin{figure}
    \begin{center}
        \begin{tabular}{|c|c|c|c|c|c|}
            \hline
                $Q$ & $\omega$ & $E$ & $B$ & $|\vec E|$ & $J$  \\
            \hline
                4* & -0.9 & \includegraphics[height=3cm]{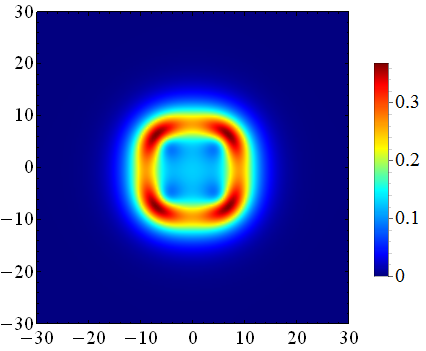} & \includegraphics[height=3cm]{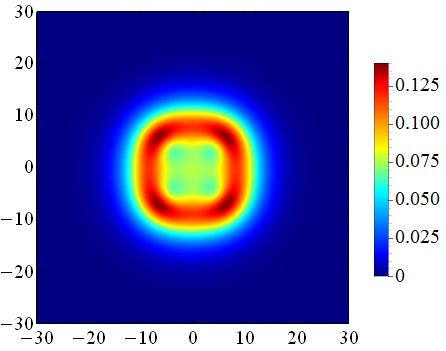} & \includegraphics[height=3cm]{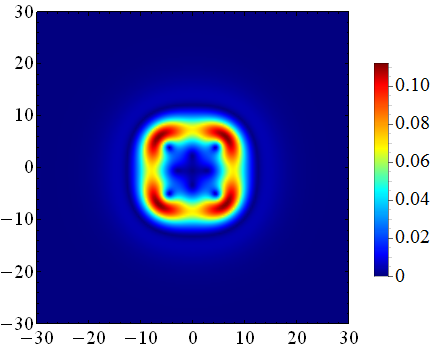} & \includegraphics[height=3cm]{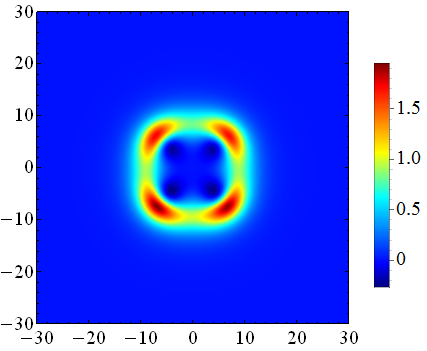} \\
            \hline
                4 & -0.9 & \includegraphics[height=3cm]{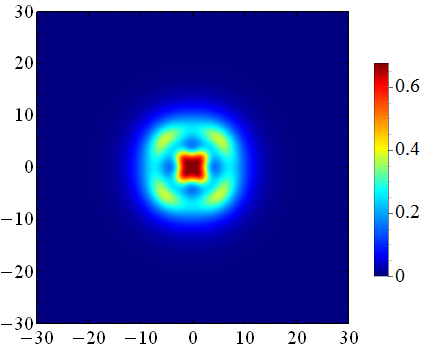} & \includegraphics[height=3cm]{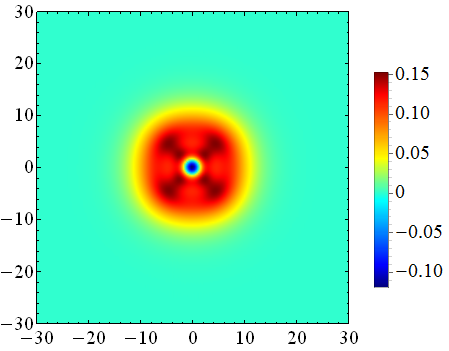} & \includegraphics[height=3cm]{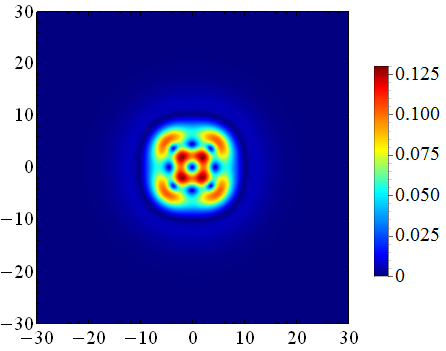} & \includegraphics[height=3cm]{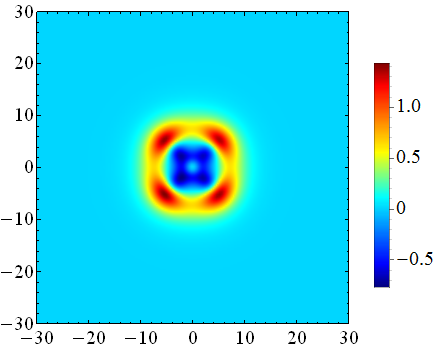} \\
            \hline
                4* & 0.9 & \includegraphics[height=3cm]{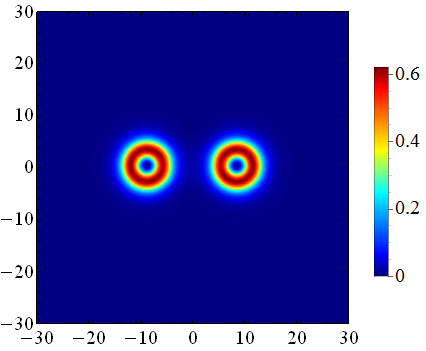} & \includegraphics[height=3cm]{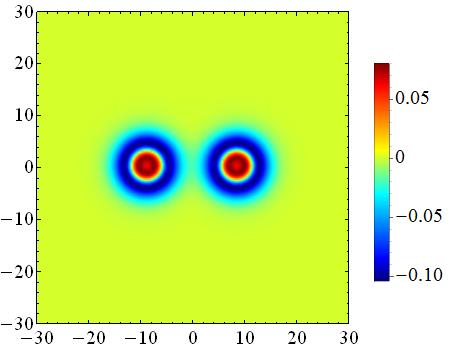} & \includegraphics[height=3cm]{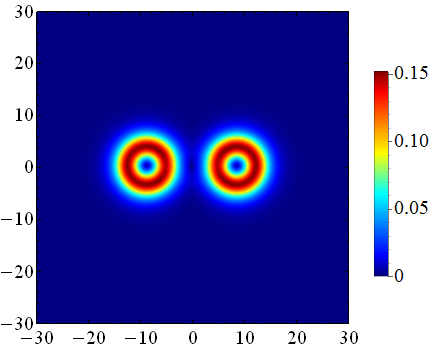} & \includegraphics[height=3cm]{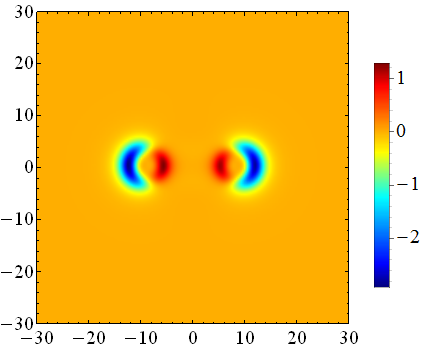} \\
            \hline
                4 & 0.9 & \includegraphics[height=3cm]{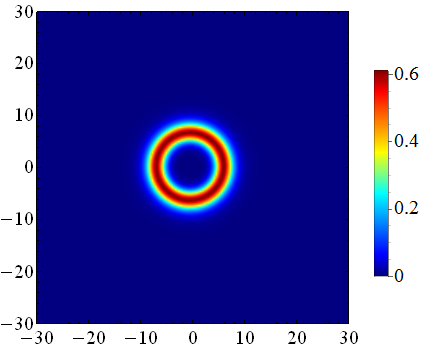} & \includegraphics[height=3cm]{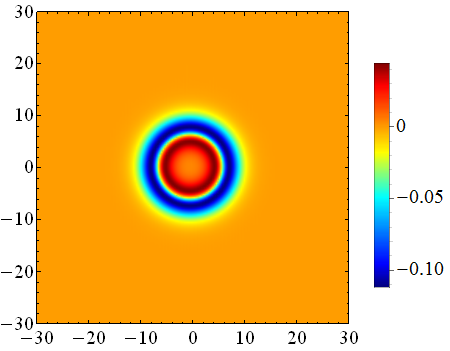} & \includegraphics[height=3cm]{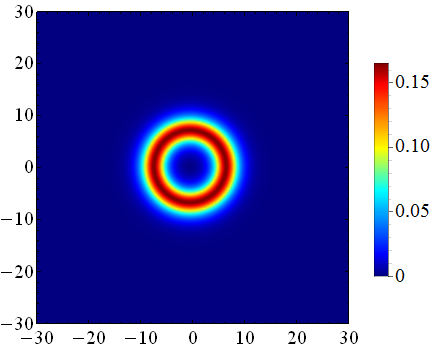} & \includegraphics[height=3cm]{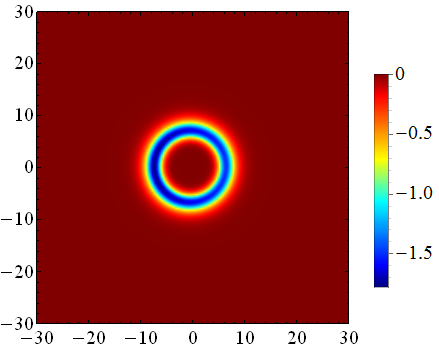} \\
            \hline
                5 & -0.9 & \includegraphics[height=3cm]{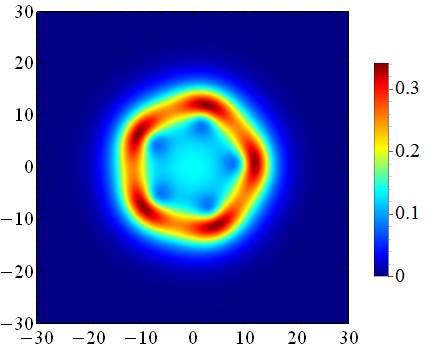} & \includegraphics[height=3cm]{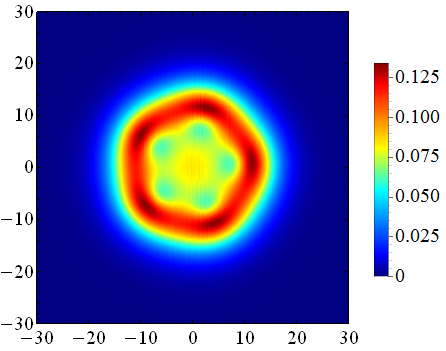} & \includegraphics[height=3cm]{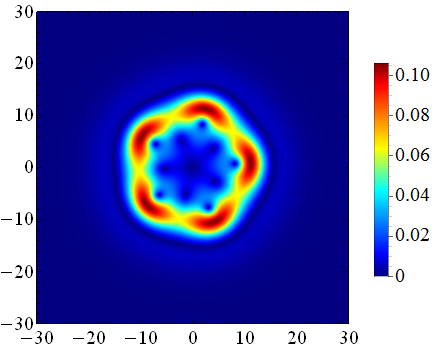} & \includegraphics[height=3cm]{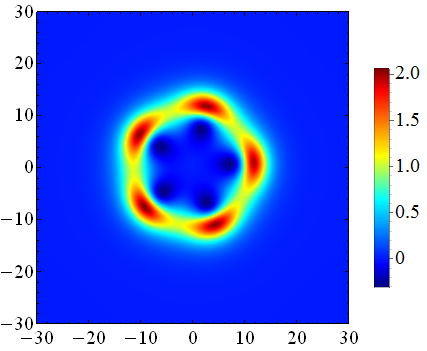} \\
            \hline
                5* & 0.9 & \includegraphics[height=3cm]{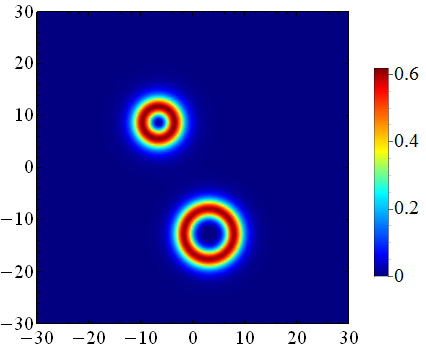} & \includegraphics[height=3cm]{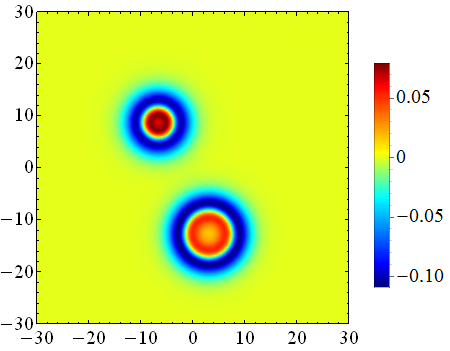} & \includegraphics[height=3cm]{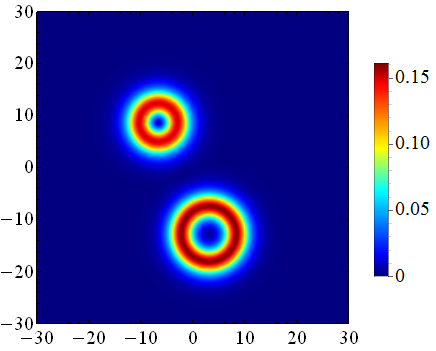} & \includegraphics[height=3cm]{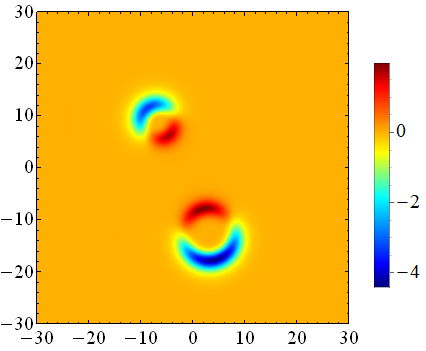} \\
            \hline
                5 & 0.9 & \includegraphics[height=3cm]{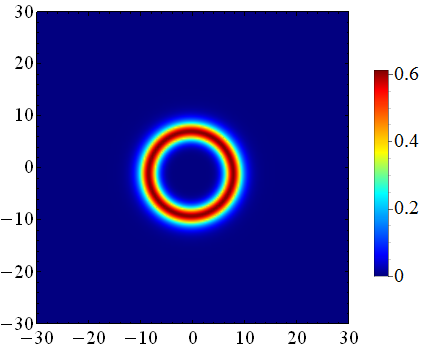} & \includegraphics[height=3cm]{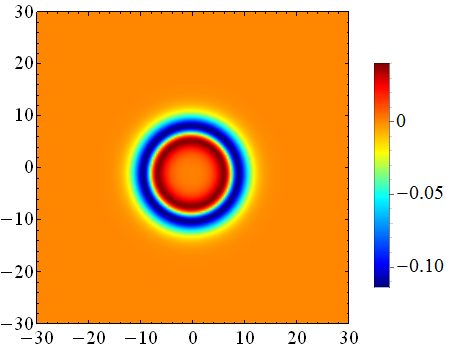} & \includegraphics[height=3cm]{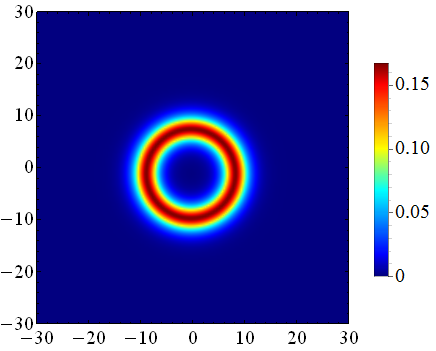} & \includegraphics[height=3cm]{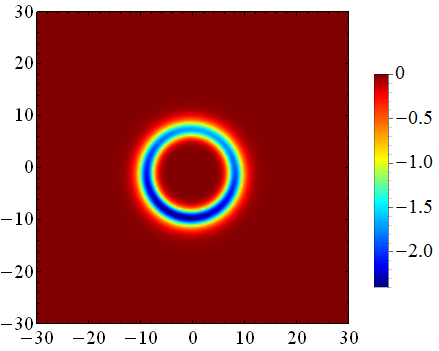} \\
            \hline
        \end{tabular}
    \end{center}
    \lbfig{table_double}
    \caption{Energy density, magnetic field, module of electric field and angular
    momentum density of solutions in model with double vacuum potential for charges $Q=1-5$, $g=0.3$ and $\omega\in[-1,1]$
    (Local minima are marked with *).}
\end{figure}

%%%%%%%%%%%%%%%%%%%%%%%%%%%%%%%%%%%%%%%%%%%%%%%%%%%%%%%%%%%%
\section{Summary}
%%%%%%%%%%%%%%%%%%%%%%%%%%%%%%%%%%%%%%%%%%%%%%%%%%%%%%%%%%%%
The objective of this work is to introduce a class of new solitons in the gauged (2+1)-dimensional
Maxwell-Chern-Simons-Skyrme model and investigate their properties.
This model can be implemented in the description of the integer quantum Hall effect, where planar Skyrmions are coupled to the magnetic
flux and possess electric charge.

We show that the electromagnetic interaction in this system may strongly affect the usual structure
of multisoliton solutions, in particular the rotational invariance
may be broken due to strong electric repulsion between the constituents. Since the usual pattern of interactions between the planar
Skyrmions depends on the particular form of the potential term, we analysed two possibilities, related with choice of the weakly
bounding potential and the double vacuum potential.

The presence of the Chern-Simons term, which breaks the time- and space-reversal symmetry,
in both cases yields the magnetic flux and electric
charge of the configuration.  In the former case, for a given particular value of the gauge coupling constant $g=0.3$ and
with the usual choice of the
parameters of the potential, the solitons remain well separated for all range of values of the electric potential.
In the latter case
the rotationally invariant configurations remain to be global minima in all sectors for all positive values of the electric potential. In the
opposite limit, as $\omega\to - 1$, the symmetry becomes broken to the dihedral group. Clearly, increase of the strength of the gauge coupling may
provide different results, in particular it may yield recovering of the rotational invariance of the solutions.
Investigation of the corresponding dependency could be an interesting study, which is,
however, beyond the purposes of this work.

Notably in the strong coupling limit both the total magnetic flux and the electric charge, associated with the solitons via the
Chern-Simons term, become effectively quantized, although they both  are not topological charges.

Finally, we found a new type of the electrically charged $Q=4$ multisolitons centered around a magnetic flux. It might be an interesting
problem to investigate which new types of solutions may appear in the sectors of higher topological degree.

%%%%%%%%%%%%%%%%%%%%%%%%%%%%%%%%%%%%%%%%%
\section*{Acknowledgements}
%%%%%%%%%%%%%%%%%%%%%%%%%%%%%%%%%%%%%%%%%
Y.S. gratefully
acknowledges support from the Russian Foundation for Basic Research
(Grant No. 16-52-12012) and DFG (Grant LE 838/12-2). He would like to thank Luis Ferreira, Derek Harland, Eugen Radu,
Paul Satcliffe and Tigran Tchrakian for useful discussions and valuable
comments. Some of the work of Y.S. was supported by the FAPESP (Grant No. 15/25779-6), he would like to thank the
Instituto de F\'{i}sica de S\~{a}o Carlos for its kind
hospitality during the completion of this work.

\end{document}